\documentclass[]{mn2e}

\usepackage{amssymb}
\usepackage{amsmath}
\usepackage{graphicx}
\usepackage{appendix}
\usepackage{longtable}
\usepackage{url}

\title[LF]{The massive end of the luminosity and stellar mass functions and clustering from CMASS to SDSS:  Evidence for and against passive evolution}
\author[Bernardi et al.]{\parbox{\textwidth}{M. Bernardi$^{1}$\thanks{E-mail: bernardm@sas.upenn.edu}, A. Meert$^{1}$, R. K. Sheth$^{1,2}$, M. Huertas-Company$^{3}$, C. Maraston$^{4}$, F. Shankar$^{5}$ \& V. Vikram$^{1}$} \vspace{0.4cm}\\
\parbox{\textwidth}{$^{1}$Department of Physics and Astronomy, University of Pennsylvania, Philadelphia, PA 19104, USA\\
$^{2}$The Abdus Salam International Center for Theoretical Physics, 
      Strada Costiera 11, 34151 Trieste, Italy\\
$^{3}$GEPI, Observatoire de Paris, CNRS, Univ. Paris Diderot;
Place Jules Janssen, 92190 Meudon, France\\
$^{4}$Institute of Cosmology and Gravitation, Dennis Sciama Building, Burnaby Road, Portsmouth PO1 3FX, UK\\
$^{5}$School of Physics and Astronomy, University of Southampton,
Southampton SO17 1BJ, UK\\}}

\begin{document}
 \date{Accepted .  Received ; in original form }

\maketitle

\label{firstpage}

\begin{abstract}
We describe the luminosity function, based on Sersic fits to the light profiles, of CMASS galaxies at $z\sim 0.55$.  Compared to previous estimates, our Sersic-based reductions imply more luminous, massive galaxies, consistent with the effects of Sersic- rather than Petrosian or de~Vaucouleur-based photometry on the Sloan Digital Sky Survey (SDSS) main galaxy sample at $z\sim 0.1$.  This implies a significant revision of the high mass end of the correlation between stellar and halo mass.  Inferences about the evolution of the luminosity and stellar mass functions depend strongly on the assumed, and uncertain, $k+e$ corrections.  In turn, these depend on the assumed age of the population.  Applying $k+e$ corrections taken from fitting the models of Maraston et al. (2009) to the colors of both SDSS and CMASS galaxies, the evolution of the luminosity and stellar mass functions appears impressively passive, provided that the fits are required to return old ages.  However, when matched in comoving number- or luminosity-density, the SDSS galaxies are less strongly clustered compared to their counterparts in CMASS.  This rules out the passive evolution scenario, and, indeed, any minor merger scenarios which preserve the rank ordering in stellar mass of the population.  Potential incompletenesses in the CMASS sample would further enhance this mismatch.  Our analysis highlights the virtue of combining clustering measurements with number counts.  
\end{abstract}


\section{Introduction}
The most massive galaxies in the Universe are interesting for several reasons, so they have been the object of much study.  Recent work has shown that the most luminous galaxies at $z\sim 0.1$ are more abundant than expected from the most commonly used parametrizations of the luminosity function (Bernardi et al. 2010, 2013).  When converted to a stellar mass function (a conversion which is rather sensitive to the assumed stellar mass-to-light ratio, about which, as we show below, one has to be very careful), this mis-match is important for models which use the observed abundance and its evolution to constrain the issue of whether these objects were assembled via major or minor mergers.  E.g., Bernardi et al. (2011a,b; 2014) and Cappellari et al. (2013) show that two mass scales are special for both early- and late-type galaxies:  $M_* \sim 3\times 10^{10}M_\odot$ and $2\times 10^{11}M_\odot$.  These scales are thought to be related to a change in the assembly histories (e.g. to ones in which wet versus dry or minor versus major mergers become important).  These mass scales are particularly interesting because it has long been thought that the most massive galaxies are also the ones whose stellar populations are most likely to have evolved passively (e.g. Cimatti et al. 2006).  If they evolve passively, or they do not merge with other members of their sample as their stellar populations age, then the fact that their comoving number density does not evolve allows one to use the evolution of their clustering strength to constrain the growth factor (e.g. Wake et al. 2008).  However, the most massive galaxies are rare, so measuring their clustering reliably requires a large volume.  

The BOSS survey (Anderson et al. 2012) defines a sample of massive galaxies -- the CMASS sample -- chosen to be a population of nearly constant comoving number density over $0.5\le z\le 0.7$.  Maraston et al. (2013; hereafter M13) showed that, across $0.4\le z\le 0.6$, the stellar masses in CMASS have evolved little if at all, especially if one restricts attention to the reddest objects (those with observed $g-i>2.35$).  Montero-Dorta et al. (2014) show that the CMASS luminosity function also appears to evolve passively over this range. One of our main goals is to explore the possibility that the CMASS galaxies evolve passively all the way to $z=0$.  This is complicated because a galaxy's luminosity can evolve significantly as its stellar population ages, and this evolution can be a function of waveband.  Both of these motivate observational estimates of the stellar mass instead, since, for a population which is neither forming new stars nor merging, this quantity (if appropriately defined) should be constant and independent of waveband.  Since stellar masses $M_*$ are determined from the product of $M_*/L$ and $L$, they tend to be less reliably determined than luminosities themselves (Bernardi et al. 2013).  To address the question of passive evolution, one must ensure that systematics associated with estimating $L$ or $M_*/L$ at different $z$ are neither hiding real, nor masquerading as, evolution.  

This is nontrivial because, even at fixed redshift, estimates of the bright end of the luminosity function depend strongly on how one fits the light profile.  Estimates based on fitting single component Sersic profile (S{\'e}rsic 1963) are much less biased than SDSS-pipeline analyses based on the de~Vaucouleur's profile or Petrosian's procedure, both of which underestimate the total light (Meert et al. 2013, 2015).  
The use of this more accurate photometry substantially increases the inferred stellar mass density at $z\sim 0.1$ (Bernardi et al. 2013).  Although this impacts studies which seek to relate the stellar mass of a galaxy to the mass of its parent halo (Kravtsov et al. 2014; Shankar et al. 2014), it makes little sense to compare their local measurement with stellar masses at $z\sim 0.55$, say, unless the high redshift sample is also based on similar photometric reductions.  In Section~\ref{sec:lf}, we use the same photometric pipeline which was used to analyze the $z\sim 0.1$ sample, {\tt PyMorph} (Meert et al. 2015), to study the CMASS sample.  

Comparison of the CMASS and SDSS luminosity and stellar mass functions requires an understanding of $k+e$ corrections.  Section~\ref{sec:passive} shows that these depend strongly on assumptions about the age of the population.  However, if we use the same models -- those of Maraston et al. (2009; hereafter M09) -- to analyze both SDSS and CMASS, {\em and} we require that the models return old ages, then passive evolution appears consistent with the data.  

There is a well-known prediction for how the clustering signal of a passively evolving population should evolve.  In view of the sensitivity to the $k+e$-corrections, in Section~\ref{sec:xi} we use the clustering of the CMASS and SDSS galaxies to check if the most massive SDSS galaxies really are simply passively evolved from CMASS.  We find that the clustering of the most massive galaxies in the SDSS is weaker than expected if the most massive SDSS galaxies were also the most massive galaxies in CMASS. Appendix~\ref{sec:toy} discusses implications of this for merger models, and a final section summarizes our findings and places them in the context of previous work.  

Where necessary, we assume a flat $\Lambda$CDM cosmology with $\Omega_{m} = 0.3$ and Hubble constant $H_0 = 70~$km~s$^{-1}$~Mpc$^{-1}$ at the present time.  Changing $\Omega_m$ by 10\% does not affect our conclusions.  All volumes and number densities we quote are comoving.  Our SDSS analysis is based on the DR7 Main Galaxy sample (Abazajian et al. 2009), which provides redshifts, {\tt cmodel} magnitudes for the apparent brightness and {\tt model} magnitudes for the colors of each object (see the survey documentation at {\tt www.sdss.org} for details of these quantities). In addition, all stellar masses we quote assume 97\% solar metallicity plus 3\% of 0.05 solar (following M13) and a Chabrier IMF.  

\section{Effect of {\tt PyMorph} photometry on the luminosity function}\label{sec:lf}

\begin{figure}
 \centering
 \includegraphics[scale = .43]{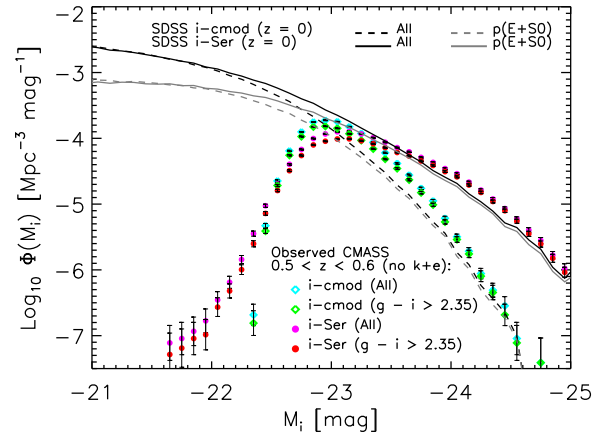}
\caption{Comparison of $i$-band luminosity functions based on {\tt cmodel} and {\tt PyMorph} {\tt Sersic} photometry.  Symbols show results for the CMASS galaxies as well as a red subsample (having observed {\tt model} $g-i>2.35$).  Solid and dashed black curves show the corresponding results for all the SDSS main sample galaxies, while gray solid and dashed curves show only the E+S0s. The reasonable agreement between CMASS and SDSS is fortuitous since no $k+e$ corrections have been applied, but this helps illustrate the fact that the difference between {\tt cmodel} and {\tt PyMorph} photometry is the same in both SDSS and CMASS.}
 \label{pymorph}
\end{figure}

In the SDSS, the bright end of the luminosity function is much brighter if {\tt PyMorph} derived Sersic rather than {\tt cmodel} photometry is used.  Our first goal is to determine if this is also true in CMASS.  The cyan and magenta symbols in Figure~\ref{pymorph} show $1/V_{\rm max}$-based estimates of the CMASS luminosity functions derived from {\tt cmodel} and Sersic photometric reductions, respectively.  Although we have used $i$-band photometry, these are not $i$-band luminosity functions in the strict sense, because no $k+e$ corrections have been applied.  This is not a concern for the present purpose, since, for each object, these would likely be the same for both photometric reductions.  The figure clearly shows that the Sersic reductions produce substantially more high luminosity objects.\footnote{Strictly speaking, Meert et al. (2013) and Bernardi et al. (2014) showed that, although single Sersic fits to galaxies at $z\sim 0.1$ are much less biased than {\tt cmodel} magnitudes, they do slightly overestimate the total light -- estimates based on fitting two component Sersic + Exponential profiles are less biased.  However, the CMASS galaxies are too distant to allow accurate determinations of the two component fits.}

For most of the remainder of this paper we will be interested in the question of passive evolution.  The reddest CMASS objects -- those with observed $g-i>2.35$ -- are much more likely to evolve passively (see, e.g., M13 for details).  Therefore, we only test for passive evolution using this redder subset.  The green and red symbols in Figure~\ref{pymorph} show the luminosity functions for this subset.  The red galaxies account for about 70 percent of all the objects at the bright end (i.e. comoving number densities less than about $0.5\times 10^{-4}$/Mpc$^3$), but otherwise the distribution of luminosities is very similar to that of the full sample.


The main point of this figure is to check if the difference between {\tt cmodel} and {\tt PyMorph} magnitudes in CMASS is similar to that in the SDSS.  This raises the question of what the analog of the $g-i>2.35$ color cut is at $z\sim 0.1$.  A simple color cut will not suffice, because it is well known that a substantial fraction of red SDSS galaxies are actually edge-on disks (i.e. are not passively evolving).  Recently, Huertas-Company et al. (2011) have used a variety of different observables (in addition to color) to assign to each galaxy a probability that it is a certain morphological type.  These Bayesian Automated Classification (hereafter BAC) probabilities are available for all the DR7 objects.  So, in what follows, we weight each galaxy in our SDSS sample by the BAC probability of Huertas-Company et al. (2011) that it is an elliptical (E) or an S0.  In support of this choice, we note that, for the luminosity thresholds we consider later, the red galaxies are the same fraction, $\sim 0.7$, of the full CMASS sample that E+S0s are of SDSS galaxies.  

The solid curves show Sersic determinations of the luminosity function of all galaxies in the SDSS and of E+S0s (obtained by weighting each SDSS galaxy by $p$(E) + $p$(S0), the probability that it is an E or S0 as determined by Huertas-Company et al. 2011).  The dashed curves show a similar analysis using {\tt cmodel} magnitudes.  Since no $k+e$ corrections are applied, the reasonable agreement at the bright end of the CMASS measurements is fortuitous.  However, this makes it easy to see that Sersic photometry results in more high luminosity objects in CMASS, quantitatively just like it does in the SDSS.

Therefore, just as the SDSS analysis results in a significant revision of the $M_*-M_{\rm halo}$ relation at $z\sim 0$, our {\tt PyMorph} reductions of the CMASS sample imply a significant revision at the high mass end of the $M_*-M_{\rm halo}$ relation at $z\sim 0.55$.  The revision is not quite as large as the figure suggests because, in the SDSS, the Sersic luminosities are biased slightly high compared to those derived from Sersic + Exponential fits (Bernardi et al. 2013; also see D'Souza et al. 2015 for a different analysis); we expect this to be true for the CMASS sample as well.  (The slight brightward bias of the Sersic values is much less than the amount by which {\tt cmodel} magnitudes are biased faintwards.)  In this context, it is worth noting that Sersic based estimates of BCGs in Sparcs and Cosmos imply a modified $M_*-M_{\rm halo}$ relation (Shankar et al. 2014) which is qualitatively consistent with what our Sersic reductions in CMASS imply.  


\section{Evidence for passive evolution from $\phi(L)$ and $\phi(M_*)$ in the SDSS and CMASS}\label{sec:passive}

The previous section showed that, when no $k+e$ corrections are applied, the SDSS and CMASS $i$-band luminosity functions are in remarkable agreement.  Meaningful conclusions about evolution rest on the accuracy of $k+e$ corrections, which we now discuss.

\subsection{Age dependence of $k+e$ corrections}
\label{sec:age}
In what follows, we will mainly use the $(k+e)$-corrections in the $r$- and $i$-bands obtained by fitting the observed CMASS colors to the single burst stellar population (hereafter SSP) models of M09, assuming almost solar metallicity (97\% solar plus 3\% of 0.05 solar, as in M13) and a Chabrier IMF (we converted from Kroupa to Chabrier IMF as described in Tab. 2 of Bernardi et al. 2010). These are the same models used by M13 for the red galaxies (i.e. those with $g-i>2.35$). (M13 used a suite of templates with different star formation histories for bluer galaxies, i.e. those with $g-i<2.35$.  Since we are only interested in the brightest galaxies which are very likely to be well-described by the passive template anyway, we do not use this suite of other templates.)  

\begin{figure}
 \centering
\includegraphics[scale = .43]{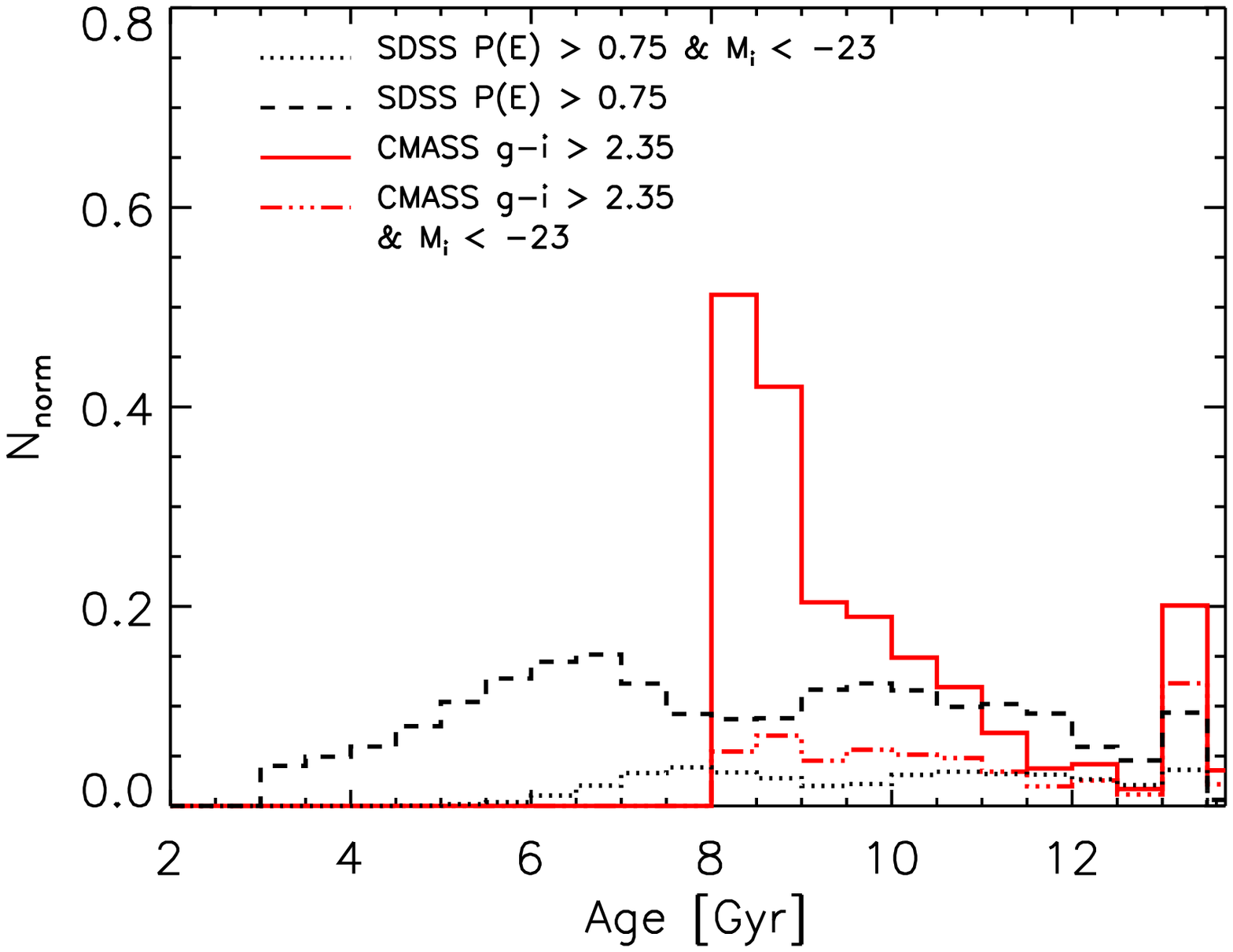}
\includegraphics[scale = .43]{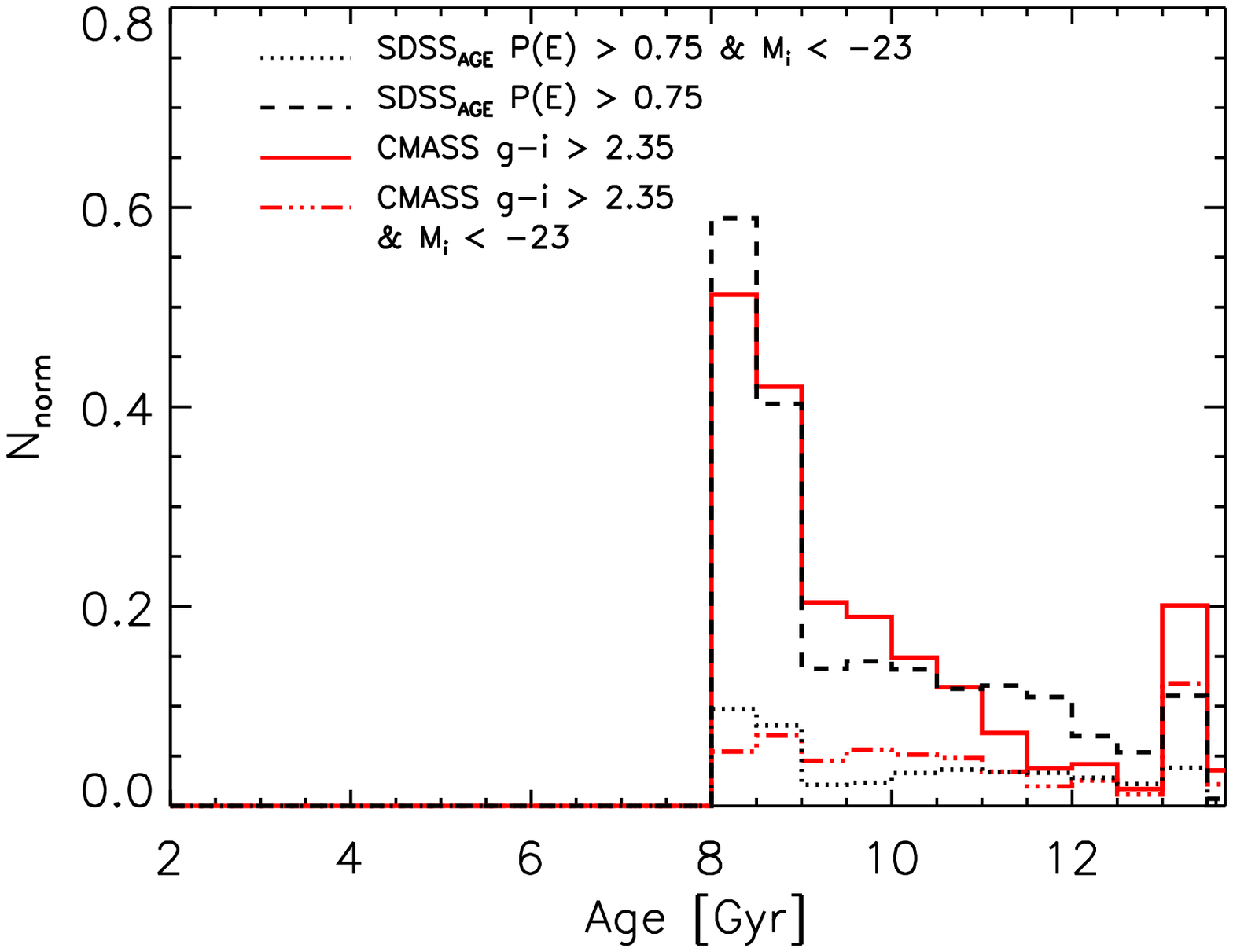}
\caption{Top panels shows the distribution of $z=0$ ages of SDSS ellipticals and CMASS red galaxies, as determined from fitting M13 models to the observed colors.  The sharp cuts at 3~Gyr and 8~Gyr, for SDSS and CMASS, are by construction.  In both cases, the less luminous galaxies tend to be younger.  SDSS ellipticals older than $\sim 5$~Gyrs may have been present at CMASS redshifts.  Bottom panel shows the age distribution of such galaxies, except that all ages between 0 and 3~Gyrs at a given $z$ have been replaced with an age of 3~Gyrs, before being shifted back to $z=0$ ages.  In effect, the distribution above 10~Gyrs is unchanged, but that between 8 and 10~Gyrs has been increased.  The age distributions, and their dependence on luminosity, are in good agreement.}
 \label{ageDist}
\end{figure}

\begin{figure}
 \centering
\includegraphics[scale = .43]{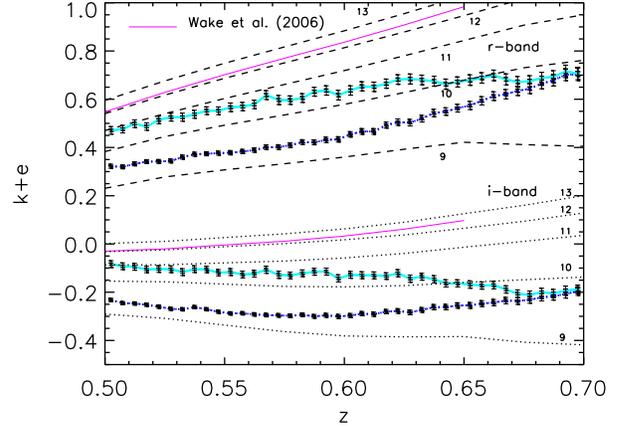}
\caption{Comparison of $(k+e)_{rr}$ and $(k+e)_{ii}$ corrections from the M09 models described in the text for a range of ages (dashed and dotted, as labelled) with those from W06 (thin solid).  Symbols with error bars show the median $k+e$ at each $z$ (plus and minus 3 times the rms error on the mean in the bin) which results from fitting the CMASS red galaxies to these M09 models; the lowest set (blue) for each band is for the full population, and the other set (cyan) is for the luminous subsample which has $M_i<-23$.  The difference indicates differential evolution of the population resulting from the fact that more luminous galaxies are older on average.  The convergence at high $z$ is spurious; it results from requiring the galaxies to be at least 3~Gyrs old at each $z$.}
 \label{keage}
\end{figure}

In practice, when M13 fit passive models, they only fit to templates with ages exceeding 3 Gyrs, but less than the age of the Universe at each $z$.  As a result, there is a lower limit on the $z=0$ age (the $z=0$ age of a redshift $z$ galaxy is the age it would have if it still exists at $z=0$), and this limit is lower for the lower redshift population.  While imposing such a lower limit is reasonable -- the CMASS colors have relatively large errors which can bias the inferred ages -- this is not the only physically reasonable choice.  For example, most previous work assumes the same $z=0$ age for all galaxies (e.g. Wake et al. 2006), so placing the same lower limit on the $z=0$ age for all galaxies would have been the most natural generalization.  

The dashed (red) and solid (black) histograms in the top panel of Figure~\ref{ageDist} show the distribution of $z=0$ ages obtained from fitting the M09 models to the red CMASS and SDSS elliptical galaxies (i.e., $p$(E)$\ge 0.75$), respectively.  The sharp cuts at 8~Gyr and 3~Gyr, for CMASS and SDSS, are by construction, as described earlier.  Most of the difference between the two distributions is due to objects which did not exist at CMASS redshifts.  To illustrate, the bottom panel shows the result of assigning each SDSS elliptical a new redshift drawn from a distribution of constant comoving number density which covers the same redshift depth as CMASS, removing from the sample all objects which were too young to have had existed at the newly assigned redshift, and replacing all ages between 0 and 3~Gyrs with an age of 3~Gyrs, before shifting back to $z=0$ ages.  This brings the two histograms into reasonably good agreement, lending qualitative support to the hypothesis that passive evolution of CMASS galaxies can account for the oldest galaxies in the SDSS.  

The dot-dot-dot-dashed (red) and dotted (black) histograms in the two panels show a similar analysis of the objects more luminous than $M_i<-23$:  they are clearly older on average.  This correlation between age and luminosity matters because the $k+e$ corrections depend strongly on age.  

Figure~\ref{keage} shows this explicitly:  the dashed and dotted lines show the $k+e$ corrections in these models for different $z=0$ ages (as labelled).  Note that these corrections are rather similar for ages between 11 and 13 Gyrs, but the age matters increasingly for younger ages.  E.g., the $k+e$ correction differs by more than 0.2~mags at $z=0.7$ for galaxies having $z=0$ ages of 9 and~10 Gyrs respectively; and $(k+e)_{rr}$ becomes negative for ages less than 8~Gyrs.  Therefore, how one imposes a lower limit to the ages when fitting to CMASS is important.

The upper most thin solid curve for each band shows the $(k+e)$-corrections used by Wake et al. (2006).  These were based on SSP models of Bruzual \& Charlot (2003), under the assumption that all the stars in a galaxy were formed in a single instantaneous burst at $z = 9.84$ (solar metallicity) after which the population evolves passively with no further star formation.  In our cosmology, this means the galaxies are assumed to be 13~Gyrs old today, and indeed, the Wake et al. (2006) $k+e$ corrections track those of M09 for this same (old) age closely.  

The lines with error bars show the median $k+e$ corrections derived for the red CMASS galaxies (i.e., $g-i>2.35$; blue lines), and for the more luminous subsample which has $M_i<-23$ (cyan lines); each error bar shows three times the rms error on the mean in the redshift bin.  Some of the tendency for the two populations to have the same $k+e$ correction at high redshift is a consequence of requiring galaxies to be at least 3~Gyrs old at the redshift of observation.  I.e., the lookback time to $z=(0.6,0.7)$ is (5.8,6.3)~Gyrs.  Hence, if a galaxy's $z=0$ age is really 9~Gyrs, at $z=0.7$ it will be (incorrectly) assigned an age of 9.3~Gyrs.  Since the $k+e$ correction is a strong function of age, particularly at younger ages, this results in a spurious upturn in $k+e$ at high-$z$.  

Notice that even the older, more luminous CMASS red galaxies lie well below the relations used by Wake et al. (2006).  Comparison with the dashed and dotted curves suggests that the full CMASS red sample is about 9 Gyrs old on average whereas the more luminous subset is about 1~Gyr older:  both are substantially younger than the Wake et al. template.  We return to this later.  The fact that the $k+e$ correction depends on luminosity is also interesting, as it is evidence of differential evolution.  This means that use of a global $k+e$ correction -- as is often assumed for massive galaxies -- may lead to biases.  Therefore, in what follows, we use the M09 corrections on an object-by-object basis.  

We have also considered $k+e$ corrections based on the Charlot \& Bruzual (2007; CB07) algorithm.  In this case, we use SSPs which have ages between 8 and 12.5~Gyrs today.  As a result, the CB07 models allow the CMASS galaxies to be slightly younger than do M13.  Provided that we shift the CB07 model fainter by 0.08~mags in $r$ before fitting, to account for known problems with the $r$-band in these models (see, e.g., M09), the $k+e$ corrections are in rather good agreement at $z\le 0.2$.  But, by $z\sim 0.6$, there are systematic differences, with the CB07 based corrections being smaller (or more negative) by about 0.1~mags. Although we are mostly interested in the $k+e$ correction, it is worth noting that the models have very similar $k$ corrections, so the differences are due to the $e$ part of the correction.  

Although we do not use the CB07-based corrections further, it is worth making the point that, without a priori knowledge of which $k+e$ correction is correct, conclusions about pure passive evolution will be limited by this uncertainty.  This is particularly worrying, since the $k+e$ corrections are very sensitive to the lower limit on the ages which has been imposed by hand.  In an attempt to determine if our decision to use corrections based on M09 is correct, we have performed two tests of the hypothesis that the CMASS galaxies have evolved passively to the present.  The first is a more careful study of the luminosity and stellar mass functions where, because of the upturn in $k+e$ at high redshifts in Figure~\ref{keage}, we confine our study to $z<0.6$.  The second uses their spatial distribution.


\subsection{SDSS$_{\rm CMASS}$:  A passively evolved mock catalog}
We begin with all the objects in the SDSS-DR7 Main Galaxy sample.  Since the SDSS is apparent magnitude limited in the $r-$band, each of these could have been observed out to a maximum comoving volume $V_{smax}$, which depends on the object's Petrosian $r$-band luminosity and the SDSS apparent magnitude limit.  We assign each object a new redshift, $z_c$, where $z_c$ is drawn from a distribution which has constant comoving density between $z=0.5$ and $z=0.6$.  We then use the same models we use to analyze the high redshift sample -- in this case, the M09 models -- to $k+e$ correct the $gri$ magnitudes of each SDSS object from their true $z$ values to their new $z_c$ values.  

\begin{figure}
 \centering
 \includegraphics[scale = .43]{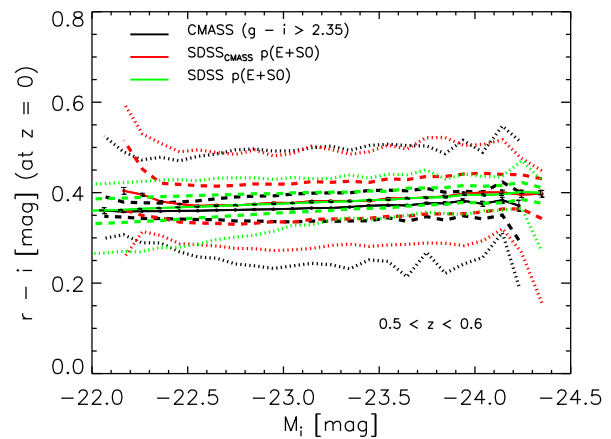}
 \caption{Absolute {\tt model}-color -- {\tt cmodel}-magnitude relation in the red CMASS $z\sim 0.55$ sample, the corresponding SDSS$_{\rm CMASS}$ sample, and the E+S0s in the SDSS, all $k+e$ corrected to $z=0$ using the M13 prescriptions.  Solid curves show the median relations, dashed and dotted curves show the region which contains 68\% and 95\% of the objects.}
 \label{Mr-Mi}
\end{figure}

Each object now has a fainter apparent magnitude, so we add (correlated) random numbers to each of the bands to mimic the noisier photometry associated with fainter apparent magnitudes.  To do this, we first measured how the error depends on band and on the type of photometry.  For {\tt cmodel} and {\tt PyMorph} a single Gaussian is sufficient.  Hence, to construct our mock catalog, to Sersic $i$-band and {\tt cmodel} $r$- and $i$-bands magnitudes we added independent Gaussian noise with rms 0.1.  For the {\tt model} magnitudes (which we use to compute colors), two Gaussian components are required, and the errors in other bands are highly correlated with those in $r$.  In this case, the rms in $(g,r,i)$ equals $(0.2,0.1,0.1)$ and the correlation coefficient between the error in $r$- and another band is 0.98.  To add the second component, we select 25\% of the objects and add Gaussian noise with rms $0.1$, again taking care to account for the correlation with the error in $r$, for which the correlation coefficient is 0.8.  

Finally, we apply the CMASS selection cuts (Anderson et al. 2012) to obtain what we call the SDSS$_{\rm CMASS}$ sample.
(This involves generating a {\tt fiber2} magnitude for each object which we estimate, following its definition, from our seeing-corrected Sersic fits.)  If our $k+e$ corrections are correct, and we have accurately accounted for the photometric errors, then weighting each SDSS$_{\rm CMASS}$ object by its $V_{smax}^{-1}$ yields a purely passively evolving population which we can compare with CMASS.  Potential tests include $\phi(L|z)$, $\phi(M_*|z)$ the distribution of colors, ages, or, as we describe later, the clustering signal.  
Since we are most interested in comparing CMASS and SDSS$_{\rm CMASS}$ to test for passive evolution, we are only interested in CMASS objects having observed $g-i>2.35$, which are much more likely to evolve passively (see M13 for details).  The corresponding cut in our SDSS$_{\rm CMASS}$ sample is to weight each object by the BAC probability $p$(E) $+$ $p$(S0) when computing quantities like $\phi(L|z)$ and $\phi(M_*|z)$.  (Recall from Paragraph 3 of Section 2 that these $p$(type) weights are always necessary because a simple color cut on the $z\sim 0.1$ population, from which we built our SDSS$_{\rm CMASS}$ sample,  will not remove red edge-on disks.  In addition, the $k+e$ corrections we used when building our SDSS$_{\rm CMASS}$ sample are not appropriate for galaxies which are not early-types; weighting by the BAC probability is a simple way of removing them from the sample, since we know they cannot possibly be the descendents of CMASS galaxies anyway. )  


\begin{figure}
 \centering
 \includegraphics[scale = .43]{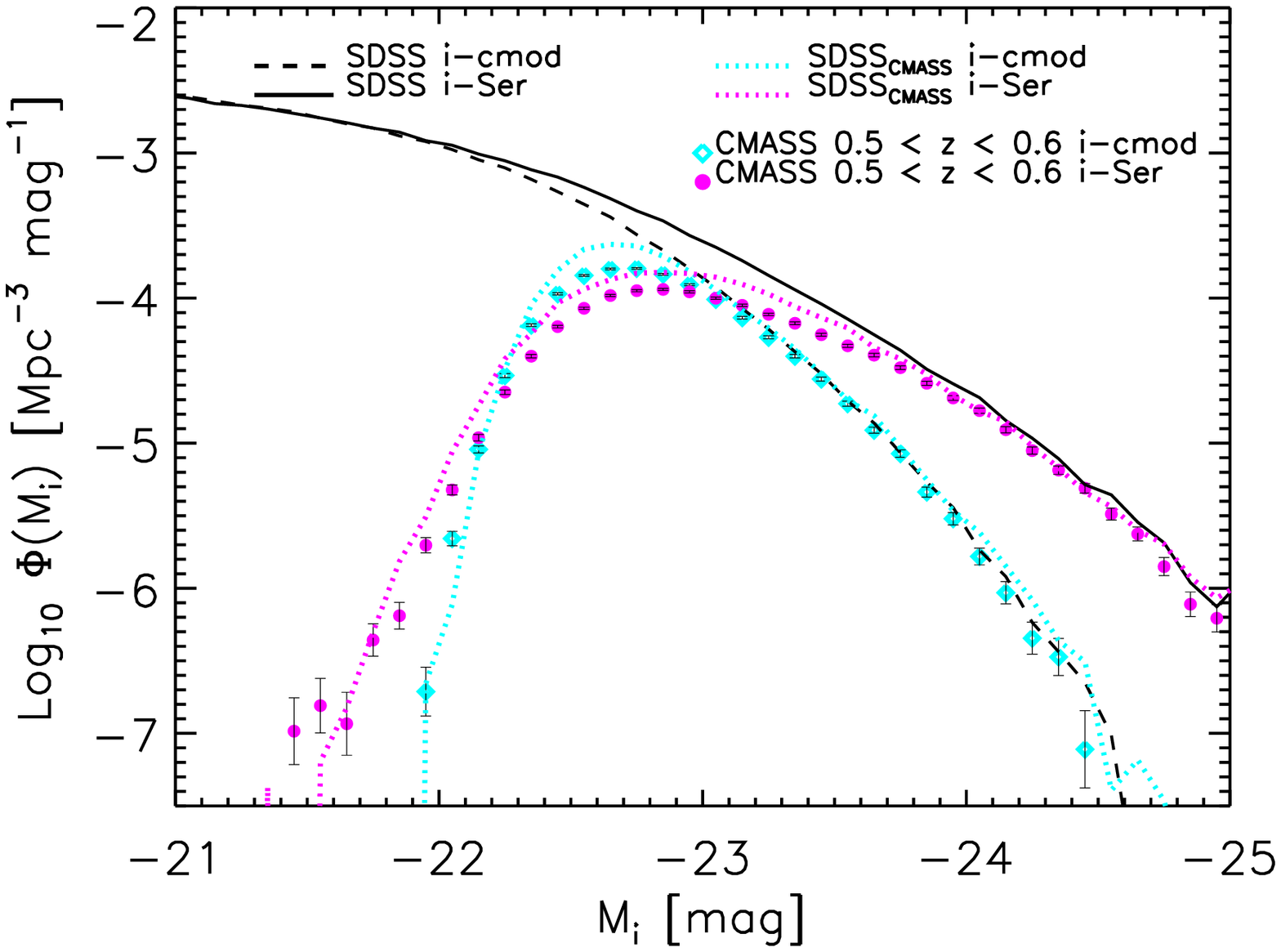}
 \includegraphics[scale = .43]{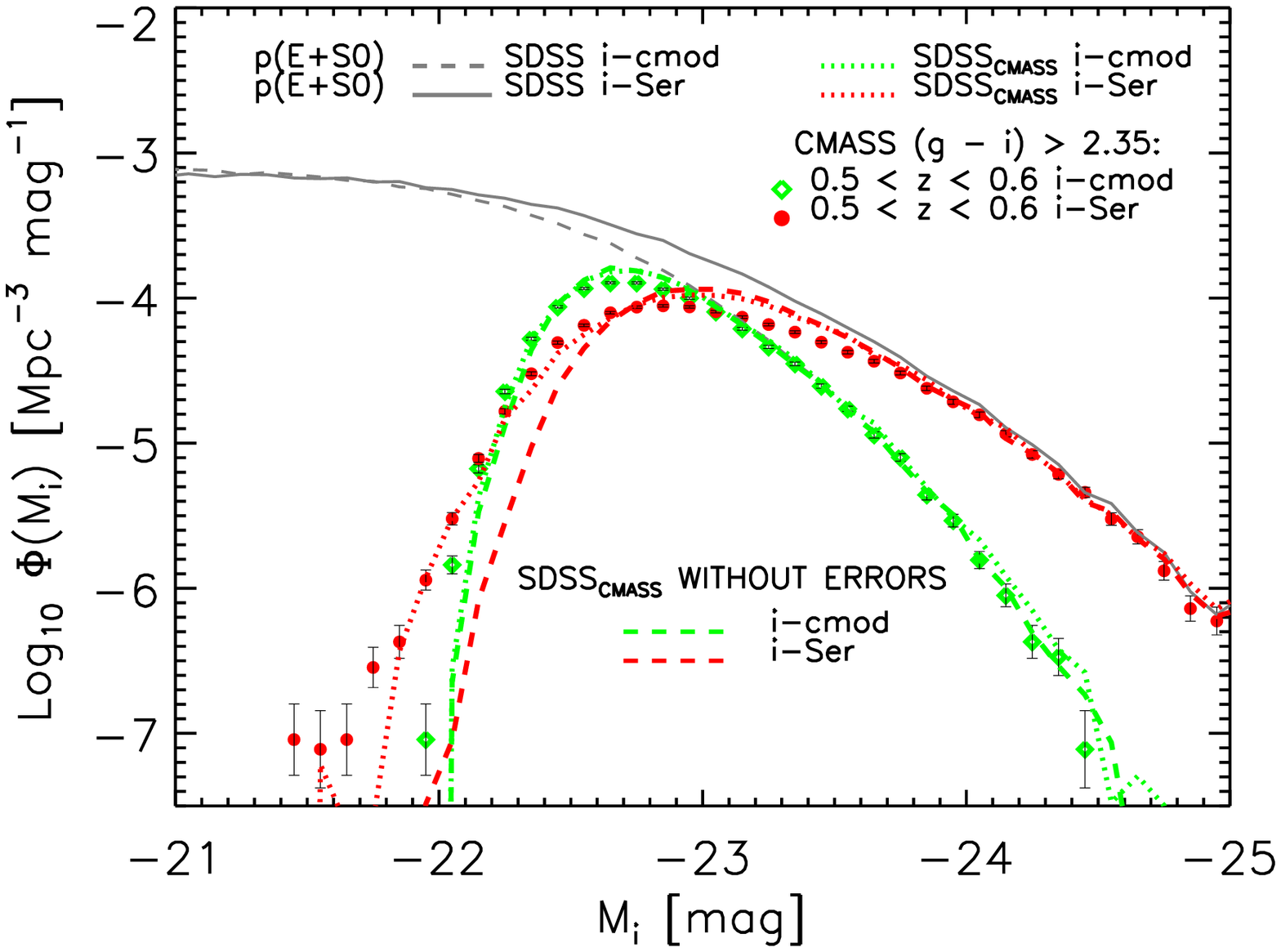}
\caption{Top panel shows the $i$-band {\tt cmodel} and {\tt PyMorph}-Sersic luminosity functions for all galaxies in CMASS (cyan and magenta symbols with error bars), SDSS$_{\rm CMASS}$ (cyan and magenta dotted lines) and SDSS (black solid or dashed lines) all corrected to $z=0$ using the M09 $(k+e)$-corrections discussed previously.  Bottom panel shows the luminosity functions for red galaxies in CMASS (red and green symbols with error bars), E+S0s in the SDSS$_{\rm CMASS}$ sample (red and green dotted curves), and E+S0s in the SDSS (gray solid and dashed) all corrected to $z=0$ using the M09 $(k+e)$-corrections.  Extra red dashed curves for SDSS$_{\rm CMASS}$ E+S0s in bottom panel show the effect of ignoring photometric errors when constructing the SDSS$_{\rm CMASS}$ mock sample.}
 \label{LFred}
\end{figure}

\subsection{Absolute magnitudes}\label{lf}


Figure~\ref{Mr-Mi} shows the color magnitude relation for the red CMASS galaxies, and the E+S0s in the SDSS$_{\rm CMASS}$ and SDSS samples, $k+e$ corrected to $z=0$.  
Dashed and dotted curves show the range in color which encloses 50 and 90 percent of the objects at each $M_i$.  The relation in the SDSS is considerably narrower than in CMASS:  this is a consequence of the larger photometric errors associated with the fainter higher redshift objects.  However, the SDSS$_{\rm CMASS}$ sample does exhibit this larger scatter, suggesting that our treatment of photometric errors is reasonably accurate.  


We now make a more careful comparison of the CMASS and SDSS$_{\rm CMASS}$ luminosity functions.  In all cases, our estimate weights each galaxy by the inverse of the comoving volume over which it could have been observed.  In principle, this comoving volume is determined by a complicated combination of the $i$-band apparent brightness and colors in the other bands.  In practice, assuming that $V_{max}$ is determined by the cut on $m_i$ only is a reasonably good approximation, so we do not include an additional term for the color cuts.  

Figure~\ref{LFred} shows the $i$-band luminosity function $(k+e)$-corrected to $z=0$ using the M09 $k+e$ corrections; the dotted curves show the {\tt cmodel} and {\tt PyMorph}-Sersic luminosity functions measured in our SDSS$_{\rm CMASS}$ samples.  These should be compared with the symbols which show the corresponding measurement in CMASS using galaxies between $0.5\le z\le 0.6$.  The top panel shows results for all galaxies, and the bottom for the subset of red CMASS galaxies (i.e., they have observed $g-i>2.35$), and E+S0s in SDSS$_{\rm CMASS}$.  The counts are in rather good agreement in the top panel, and in even better agreement in the bottom.  The fact that our SDSS$_{\rm CMASS}$ counts lie slightly but consistently above the CMASS counts may be indicating incompleteness in CMASS.  The required offset is a factor of 1.25, which is in good agreement with the recent estimate of 20 percent incompleteness by Leauthaud et al. (2015).  However, this conclusion rests heavily on the assumption that our $k+e$ corrections are indeed realistic, and that our treatment of the photometric errors is as well.  

Before we discuss these, notice that the counts are also in good agreement with the bright end of the dashed and solid gray curves.  These show the {\tt cmodel} and {\tt PyMorph}-Sersic luminosity functions in the full SDSS sample (weighted by $p$(E) $+$ $p$(S0), of course), respectively.  The dotted (SDSS$_{\rm CMASS}$-based) curves overlap the SDSS measurements at the bright end, but fall off rapidly at the faint end.  This fall-off is the expected consequence of the CMASS-selection cuts.  However, the exquisite match at the bright-end indicates that the bright end of the SDSS$_{\rm CMASS}$ sample is made up of the brightest galaxies in SDSS.  Therefore, the remarkable agreement between CMASS, SDSS$_{\rm CMASS}$ and SDSS in Figure~\ref{LFred} -- for both {\tt cmodel} and {\tt PyMorph} photometry -- suggests that the red CMASS sample is related to the (bright-end of) the SDSS E+S0 sample by purely passive evolution.  Of course, this agreement depends on the choice of $k+e$ correction, and our treatment of photometric errors.  

To show that these matter, the red and green dashed curves in the bottom panel show the result of ignoring these errors when constructing the SDSS$_{\rm CMASS}$ sample.  Clearly, this matters much more for the faint end, and more for Sersic photometry.  The dependence on photometry is not because {\tt PyMorph} photometry is noisier than {\tt cmodel}.  Rather, it is more closely related to the fact that sample selection is done using {\tt cmodel} photometry.  See Appendix~A for a more detailed discussion.

\subsection{Stellar masses}
We expect the differences in $\phi(L)$ to carry over to the stellar mass functions.  However, as Bernardi et al. (2013) have highlighted recently, this is not entirely straightforward because $M_*$ is estimated from the product of $M_*/L$ and $L$.  Therefore, $M_*/L$ can be systematically different even when $L$ is the same, or systematic differences in $L$ may not be not balanced by differences in $M_*/L$.  

\begin{figure}
 \centering
 \includegraphics[scale = .43]{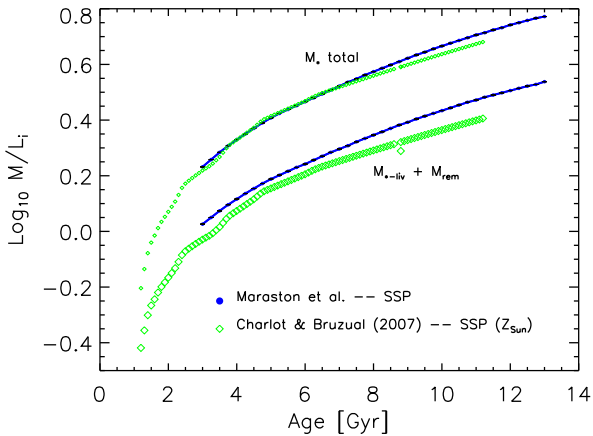}
\caption{Stellar mass to light ratios as a function of age output by the CB07 (green diamonds) and M09 (blue asterisks) models for solar metallicity and a Chabrier IMF.  Bottom curves account only for the mass in stars and stellar remnants; top curves include the mass in processed gas as well. These differences directly impact the transformation from $\phi(L)$ to $\phi(M_*)$.}
 \label{compareML}
\end{figure}

To illustrate the level at which systematics matter, Figure~\ref{compareML} shows how $M_*/L$ depends on the age of the stellar population in the CB07 and M09 SSP models we used to estimate the $(k+e)$-corrections.  The bottom blue asterisks show $M_*/L$ in the M09 models when $M_*$ is the mass in stars $M_{\rm liv}$ and stellar remnants $M_{\rm rem}$; the top blue asterisks add the gas which has been returned to the ISM by stellar evolution to give $M_{\rm tot}$.  

This raises the question of which $M_*/L$ estimates one should use?  There are two reasons why one might use the values based on $M_{\rm tot}$.  One is that $M_{\rm tot}$ is constant whereas $M_{\rm liv}+M_{\rm rem}$ evolves (as M13 note, the stellar mass lost to stellar evolution can be large:  $\sim 40$~percent at $z\sim 0.5$).   Hence, not accounting for the mass in gas compromises one of the advantages of stellar mass relative to luminosity when testing for passive evolution.  The second reason to prefer the total $M_*/L$ values is that they are in better agreement across the models:  the M09 and CB07 models differ more in the amount of gas returned to the ISM than in the total mass-to-light ratio.  
Since passive evolution conserves $M_{\rm tot}$ and not the other quantity anyway, Figure~\ref{compareML} argues that $M_*/L$ measurements in the future should be based on $M_{\rm tot}$.  

In practice, we are primarily interested in older galaxies, for which stellar evolution has slowed substantially, so the assumption that the mass in stars is constant between $z\sim 0.6$ and $z\sim 0$ is plausible.  Thus, the potential advantage of using $M_{\rm tot}$ is not so great.  In addition, the difference between the model predictions for $M_{\rm liv}+M_{\rm rem}$ is typically less than 0.1~dex.  Therefore, in what follows we will follow M13 in working with the mass currently in stars, rather than the total mass ever in stars.  But the difference between the two should be borne in mind.  



\begin{figure}
 \centering
 \includegraphics[scale = .43]{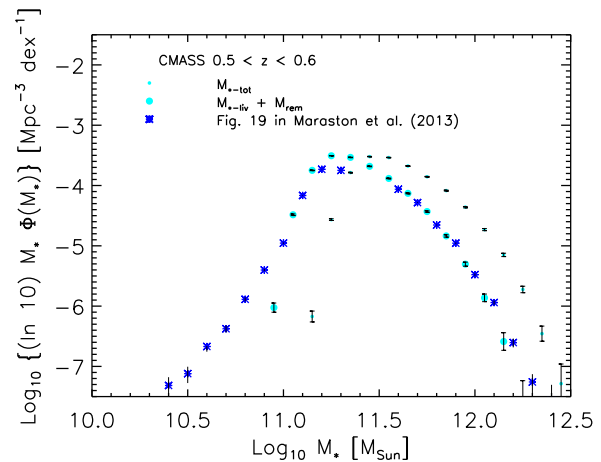}
 \caption{The CMASS stellar mass function based on luminosities derived from {\tt cmodel} $i$-band photometry.  Large cyan symbols connected by a solid line count the mass in stars and stellar remnants; small cyan symbols include the mass in gas as well, and blue symbols are taken from Figure~19 in M13.  }
 \label{M*maraston}
\end{figure}

The large cyan symbols in Figure~\ref{M*maraston} show the result of starting from the same SDSS {\tt cmodel} luminosities used by M13 (those which led to the cyan symbols in the top panel of Figure~\ref{LFred}) and using the SSP models of M09 to compute the $M_*/L$ values from which to determine $M_*\equiv M_{\rm liv}+M_{\rm rem}$.  As a consistency check, the blue symbols show the number counts reported in Figure~19 of M13; at large stellar masses they are very similar to our larger cyan symbols, as they should, since most of these massive galaxies have $g-i>2.35$.  The difference at smaller masses is due to the fact that M13 used a suite of templates with different star formation histories for bluer galaxies (i.e. those with $g-i<2.35$) whereas we did not (as we noted earlier).  The smaller cyan symbols in Figure~\ref{M*maraston} show the result of including the mass that is now in the form of gas for the M09 models; this results in an increase by a factor of 1.5.

\begin{figure}
 \centering
 \includegraphics[scale = .43]{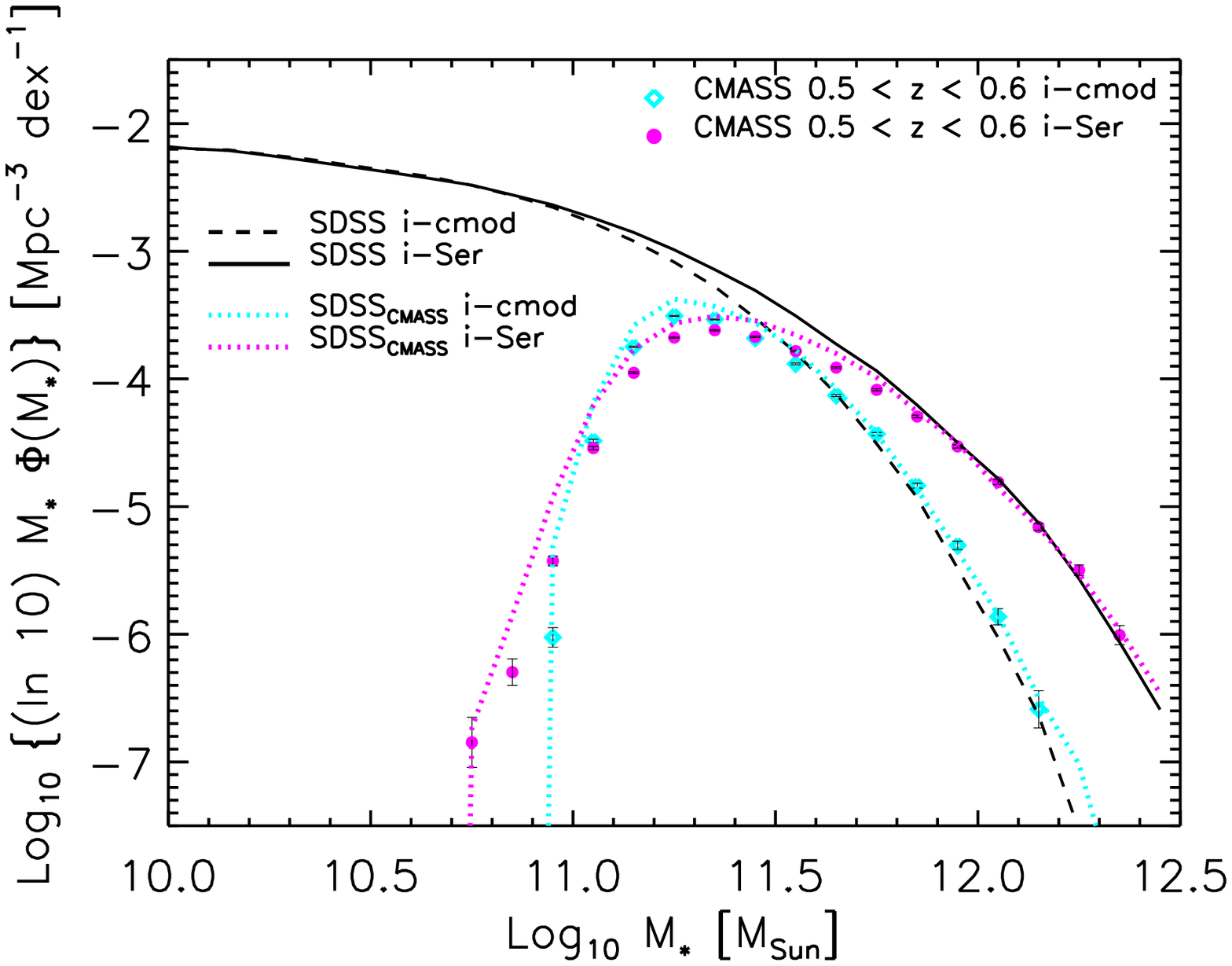}
 \includegraphics[scale = .43]{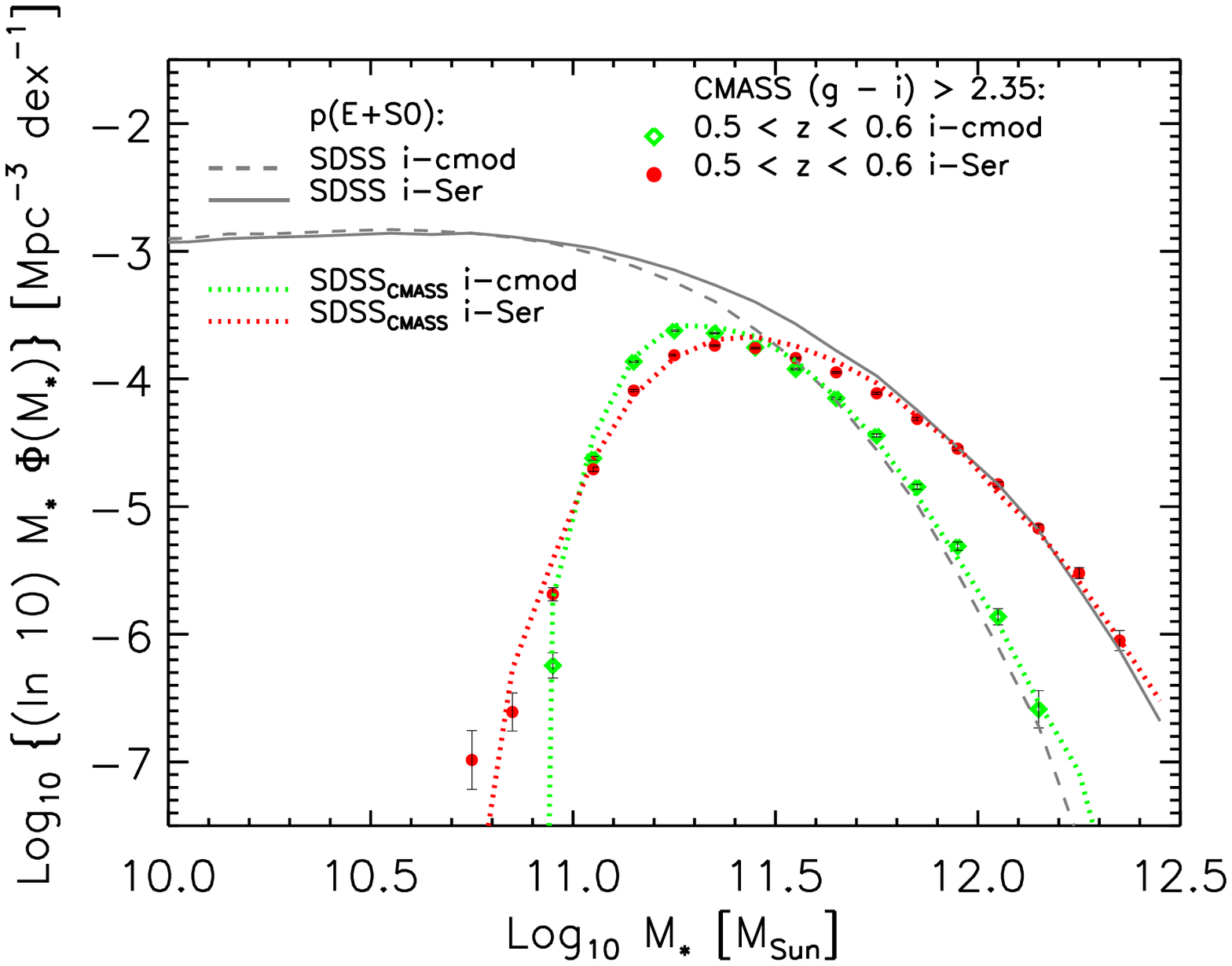}
 \caption{Comparison of stellar mass functions of all (top) and red (bottom) galaxies in CMASS with all and E+S0 galaxies in SDSS$_{\rm CMASS}$ and SDSS.  In all cases, $M_*$ comes from combining the M09 models with {\tt cmodel} or {\tt PyMorph} photometry.}
 \label{MFred}
\end{figure}


Figure~\ref{MFred} shows the $\phi(M_*)$ estimates which result from combining the luminosities which led to Figure~\ref{LFred} with $M_*/L$ estimates from the same M09 models (ignoring the mass in gas).  The top panel shows $\phi(M_*)$ for all galaxies in CMASS, SDSS, and SDSS$_{\rm CMASS}$, and the bottom panel for the reds in CMASS and the E+S0s in SDSS and SDSS$_{\rm CMASS}$.  The agreement between CMASS and SDSS$_{\rm CMASS}$, which is already good in the top panel, is even better in the bottom.  Passive evolution between CMASS and the SDSS appears to be an excellent approximation.   

Note that using the same expression for converting from $L$ to $M_*$ was crucial.  Had we used the M09 $M_*/L$ to estimate CMASS $M_*$ values, but one of the relations from Bernardi et al. (2010) to estimate SDSS$_{\rm CMASS}$ values, then we would have found that CMASS galaxies were more massive than the most SDSS$_{\rm CMASS}$ massive galaxies.  Indeed, this was the puzzle raised by M13:  How can the high redshift sample be more massive?  They noted that it was possible that systematically different $M_*$ estimates might be the reason, and our analysis appears to confirm this.  (Figure~\ref{LFri} and associated discussion argues that measurement errors do not affect this conclusion about the bright-end.)  




Although we do not show it here, analysis of the higher redshift range $0.6\le z\le 0.7$ yields similar results.  We have highlighed the $0.5\le z\le 0.6$ range because biases resulting from the M13 age requirement are less of an issue at low redshifts (c.f. Figure~\ref{keage} and associated discussion).  

The agreement between CMASS and SDSS$_{\rm CMASS}$ is consistent with passive evolution.  However, this conclusion depends crucially on the accuracy of the $k+e$ corrections (and our treatment of the photometric errors).  For this reason, we now turn to a very different test of the passive evolution hypothesis.

\section{Comparison of clustering in SDSS and CMASS}\label{sec:xi}
In this section, we use the clustering of the CMASS and SDSS$_{\rm CMASS}$ samples to determine if the two are simply related by passive evolution.  Similar tests are described in Wake et al. (2008), Tojeiro et al. (2012) and Guo et al. (2013).  

Note that it is conventional in the literature on large scale structure to work in units in which $H_0 = 100h$~km~s$^{-1}$Mpc$^{-1}$, so all distances are quoted in units of $h^{-1}$Mpc.  We will do so here, but remind the reader that the number densities in the previous section all use $h=0.7$.  In addition, whereas the previous section restricted attention to $z<0.6$ (because of potential systematics in the $k+e$ corrections), here we also include galaxies at higher redshifts.  This is because -- as we describe below -- for this test we care mainly about the rank ordering of the luminosities (or stellar masses) than their absolute values, and we do not expect the potential systematic in $k+e$ corrections to change the rank order.  

\subsection{Clustering of conserved tracers}
If CMASS galaxies evolve passively, then their comoving number density will remain unchanged.  In this case, their large scale clustering signal should evolve as 
\begin{equation}
 \xi(r|z) = b_z^2\,\xi_m(r|z) = [D_z + b_0-1]^2\,\xi_m(r|z=0)
 \label{xirz}
\end{equation}
where $D_z$ is the linear theory growth factor at redshift $z$ in units of its value at $z=0$, and $b_z$ is the bias of the population at redshift $z$ (Nusser \& Davis 1994; Mo \& White 1996).  Therefore, 
\begin{equation}
 \frac{\xi(r|z)}{\xi(r|0)} = \frac{[D_z + b_0-1]^2}{b_0^2}
                           = \frac{b_z^2}{[D_z^{-1} + b_z-1]^2}.
 \label{xizratio}
\end{equation}
(The growth on smaller scales can be slightly different; see Wake et al. 2008 for some explicit examples of the expected magnitude of this difference.)  If we measure the redshift space distorted signal $\xi(s)$, then on the scales where equation~(\ref{xizratio}) applies we expect 
\begin{equation}
 \frac{\xi(s|0)}{\xi(s|z)} \approx 
  \frac{\left(b_0^2 + 2b_0 f_0/3 + f_0^2/5\right)}
       {\left(b_z^2 + 2b_z f_z/3 + f_z^2/5\right)\,D_z^2}
 \label{xisratio}
\end{equation}
where $f_z\approx \Omega_m(z)^{5/9}$ (Kaiser 1987).  

Although we introduced equations~(\ref{xirz}--\ref{xisratio}) in the context of passively evolving galaxies, they are more generally applicable to any population of tracers having conserved comoving density.  E.g., suppose that CMASS galaxies only merge with non-CMASS galaxies.  Then their luminosities and stellar masses will almost certainly be inconsistent with those of passive evolution models.  However, their comoving number density will remain unchanged.  If these differences from true passive evolution preserve the rank order -- the most luminous/massive CMASS galaxy remains the most luminous/massive one at lower redshift -- then the clustering at fixed comoving number density (not fixed stellar mass or luminosity!) will obey equations~(\ref{xirz}--\ref{xisratio}).  Another example, which is potentially relevant to the discussion of the previous section, is to suppose that SSP models differ from one another only in the strength of the evolution in luminosity or stellar mass, such that although they predict different luminosities, they all have the same rank ordering.  In this case also, the clustering at fixed comoving number density should obey equations~(\ref{xirz}--\ref{xisratio}).  

Previous work (Anderson et al. 2012; Nuza et al. 2013; Guo et al. 2013) has shown that $b_z\approx 2$ for the full CMASS sample.  Hence, equation~(\ref{xirz}) indicates that the low redshift sample should be more strongly clustered if the CMASS galaxies have survived to become the SDSS$_{\rm CMASS}$ galaxies because of passive evolution (as Figures~\ref{LFred} and~\ref{MFred} suggest).
In our background cosmology $D_{0.55} = 0.75$, $D_{0.2} = 0.9$, $f_{0.55} = 0.76$, $f_{0.2}=0.62$ and $f_0=0.51$, so if $b_z\sim 2$ at $z\sim 0.55$, then $\xi(r|0.55)/\xi(r|0) = 0.73$ and $\xi(s|0.55)/\xi(s|0) = 0.78$, whereas $\xi(r|0.55)/\xi(r|0.2) = 0.83$ and $\xi(s|0.55)/\xi(s|0.2) = 0.85$.  
Therefore, if the objects in our SDSS$_{\rm CMASS}$ sample are descendents of those in the CMASS sample, then, at constant comoving number density, we expect the clustering amplitude to have increased by about 20 percent.  (Current uncertainty about the background cosmology means this number is uncertain by only a few percent.  This number is about three times larger than the fractional change in the growth factor over the redshift range spanned by CMASS.)  
On the other hand, if there were some CMASS-CMASS mergers, then the clustering signal will evolve differently.  The CMASS and SDSS$_{\rm CMASS}$ samples are large enough that this signal of passive-like evolution should be detectable.  

Guo et al. (2013) have performed a test of the passive-like/conserved tracer evolution hypothesis over the redshift range covered by CMASS (approximately 0.45-0.65).  Their Fig.~13 shows their results, which they argue are consistent with passive evolution.  However, because the expected fractional change in the clustering strength over the CMASS redshift range is only of order ten percent, their measurements are also consistent with no evolution whatsoever.  Indeed, Reid et al. (2014) use this lack of evolution to argue that one can simply ignore evolution across the entire CMASS sample.  

Some of the mildness of the measured evolution across the CMASS sample can be attributed to the joint effects of passive-like evolution and luminosity- or type-dependent clustering in a survey in which the observed mix of galaxy types depends on redshift.  This is relevant because Guo et al. (2013) have shown that clustering in CMASS depends on luminosity, and the fainter CMASS galaxies are only observed at the low redshift end of the sample.  Let $f$ denote the fraction of faint galaxies, $b$ the bias of this faint subset, and $B$ that of the brighter objects.  At the high redshift end, we only observe the bright objects:  suppose their clustering signal is $B_{hi-z}^2\xi_{hi-z}$.  If these bright objects evolve passively, then at lower redshifts their clustering signal is $B_{lo-z}^2\xi_{lo-z} = (D_{lo-z} + B_0-1)^2\,\xi_0$.  Although these objects accounted for the full observed population at high-$z$, they only account for $1-f$ of the observed objects at lower $z$.  The clustering of the full low redshift sample is $(fb_{lo-z} + (1-f)B_{lo-z})^2\xi_{lo-z}$.  The high and low redshift samples will have the same observed clustering signal if 
 $fD_{lo-z}b_{lo-z} + (1-f)D_{lo-z}B_{lo-z} = D_{hi-z}B_{hi-z}$.  A little algebra shows this occurs if 
 $f (B_{lo-z} - b_{lo-z}) = 1 - D_{hi-z}/D_{lo-z}$.  Since the right hand size is positive, and $f<1$, this will be satisfied only if $(B_{lo-z} - b_{lo-z})$ is large enough.  
If it is, then the luminosity dependence of clustering can cancel the effect of passive evolution in a survey (such as CMASS) in which the fainter galaxies are not seen at the highest redshifts.  To ensure that we are not affected by this, we will work exclusively with volume limited samples.

\subsection{Technical note}
In practice, the clustering test is complicated by the fact that equation~(\ref{xisratio}) is only expected to apply on scales of order $10h^{-1}$Mpc or larger.  Since most of the signal comes from smaller scales where this expression may not be extremely accurate, and we care about ten percent effects, it is desirable to test equation~(\ref{xizratio}) directly, rather than via~(\ref{xisratio}).  Following Davis \& Peebles (1983) it is commonly assumed that this can be done by measuring the projected quantity $w_p(r_p)$, which, in principle, is not affected by redshift space distortions.  However, although the definition of $w_p(r_p)$ involves an integral over all pair separations along the line of sight, the measurement is usually restricted to pair separations smaller than some scale that is typically of order $60h^{-1}$~Mpc.  As a result, the measured quantity is not completely independendent of redshift space distortions, and, if one is interested in ten percent level effects, this matters.  

For example, Reid et al. (2014) present measurements of both $\xi(s)$ and $w_p(r_p)$ for the full CMASS sample (but not for the red subset of most interest to us!).  Although they do not say so, the usual naive interpretation of the two measurements returns estimates of the large scale bias factor which differ by more than ten percent:  the bias inferred from fitting $w_p(r_p)$, when inserted into Kaiser's formula results in an overestimate of $\xi(s)$ of more than twenty percent.  Unfortunately, this systematic is precisely the expected magnitude of the passive evolution signal.  

Although this drawback of $w_p(r_p)$ has been known since its inception, it is only recently that datasets have reached the precision where this matters.  van den Bosch et al. (2013) describe how to modify the estimator of $w_p(r_p)$ to mitigate this effect.  They estimate a multiplicative correction factor which their Figure~5 suggests is approximately (i.e., to within a few percent) independent of galaxy type.  Therefore, although measurements of $w_p(r_p)$ in two samples may each be biased, their ratio is not.  We make use of this below.  

\begin{figure}
 \centering
 \includegraphics[scale = .4]{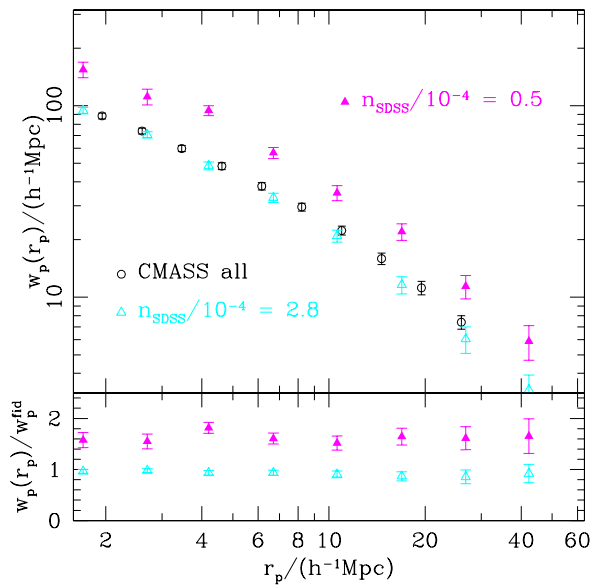}
 \caption{Projected two point correlation function of all CMASS galaxies (open circles, from Table~2 in Reid et al. 2014) and of the most luminous SDSS galaxies, selected to have comoving number densities as labelled (triangles, taken from Table C7 of Zehavi et al. 2011).  Bottom panel shows the SDSS measurements divided by $w_p$ of the full CMASS sample.  }  
 \label{xi79}
\end{figure}

\subsection{Clustering of the most luminous objects in CMASS and SDSS$_{\rm CMASS}$}

To set the stage, we first compare measurements of $w_p(r_p)$ taken from the literature.  The open circles in Figure~\ref{xi79} show the values for the full CMASS sample taken from Table~2 of Reid et al. (2014).  While they do not quote a comoving number density for their sample, their measurements are almost indistinguishable from those of Nuza et al. (2013), who quote $n = 3.6\times 10^{-4}h^3$Mpc$^{-3}$.  The SDSS DR7 sample with the most similar clustering signal has $M_r<-21.5$ and $n = 2.8\times 10^{-4}h^3$Mpc$^{-3}$; the triangles show $w_p$ for this sample taken from Table C7 of Zehavi et al. (2011).  

Before we discuss the relative amplitudes, it is worth noting how remarkably similar the shapes of curves are:  the bottom panel shows that they differ by a single scale independent multiplicative bias factor.  While there certainly are pairs common to the two SDSS samples, there are none in common with CMASS, so the agreement in shape is truly remarkable.

We now discuss the amplitudes, bearing in mind that we are most interested in comparisons at fixed comoving number density.  We use the results of Nuza et al. (2013) to account for the fact that the CMASS abundances are larger.  On the basis of mock catalogs matched to CMASS they report that the clustering strength increases as the abundance decreases; a CMASS sample with $n = 2.8\times 10^{-4}h^3$Mpc$^{-3}$ would be 6\% more strongly clustered than we have shown.  This would make the SDSS sample slightly less strongly clustered than a CMASS of the same number density.  This conflicts with the conserved tracers prediction that SDSS should be of order 20\% more strongly clustered.  

The discrepancy may not be unexpected, since one expects passive evolution to be a better approximation for the rarer more massive objects.  The filled triangles show an SDSS DR7 sample having $n = 0.5\times 10^{-4}h^3$Mpc$^{-3}$ (again from Table C7 of Zehavi et al. 2011).  The Nuza et al. scaling of $b$ with $n$ indicates that the clustering should be 60\% higher.  This is similar to or slightly larger than the corresponding SDSS measurements, so also conflicts with the conserved tracers prediction.

Since it is possible that this discrepancy is due to contamination by bluer galaxies in both CMASS and SDSS, we made our own measurements of red galaxies in CMASS and E+S0s in the SDSS.  To check consistency of our measurement software with previous work, we first measured $\xi(s)$ and $w_p(r_p)$ in the full CMASS sample, using the weighting scheme detailed in Anderson et al. (2012), finding good agreement with $\xi_{\rm N+S}$ of Table~2 in Nuza et al. (2013) and with Tables~2 and~3 in Reid et al. (2014).\footnote{However, in what follows, we work with volume limited catalogs, so the $w_{\rm FKP}$ weights of all the objects in a given catalog are the same.  Also, we are only interested in scales on which fiber collisions matter at less than the few percent level.}  

\begin{figure}
 \centering
 \includegraphics[scale = .4]{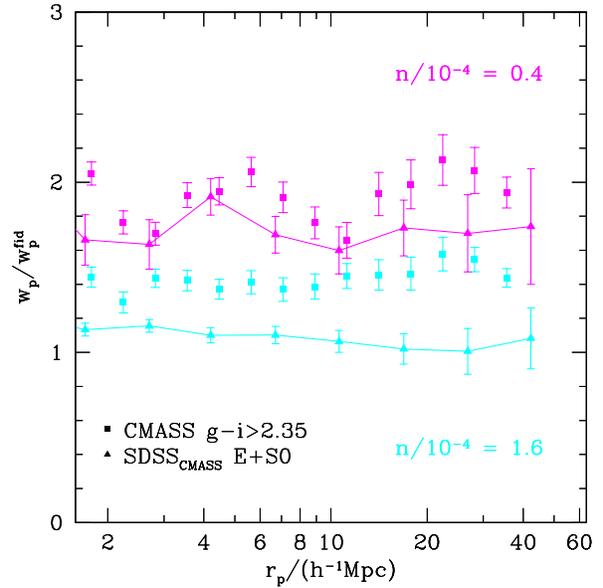}
 \caption{Ratio of projected two point correlation function of CMASS red galaxies (squares) and SDSS$_{\rm CMASS}$ E+S0 galaxies (triangles, connected by lines) to that of the full CMASS sample (open circles in previous figure).  Upper symbols show results for comoving number densities of $0.4\times 10^{-4}h^3{\rm Mpc}^{-3}$; lower symbols are for $1.6\times 10^{-4}h^3{\rm Mpc}^{-3}$.  If the conserved tracer/passive-like hypothesis were correct, then the SDSS$_{\rm CMASS}$ galaxies would be about 20\% more strongly clustered than their CMASS counterparts.}
 \label{compareWp}
\end{figure}

Having established that our software reproduces published results, we next made three volume limited catalogs, defined by choosing the most luminous CMASS galaxies having redshifts in the range $z_{min}=0.5$ and $z_{max} = (0.62,0.67,0.67)$; the associated luminosity thresholds $M_i<(-22.62,-22.87,-23.08)$ are chosen to yield comoving number densities of $n = (2,1,1/2)$ $\times 10^{-4}h^3$Mpc$^{-3}$.  
We then measured $\xi(s)$ and $w_p(r_p)$ in each of these samples and found that the more luminous samples were more strongly clustered.  This extends the findings of Guo et al. (2013) to higher luminosities.  Since this is not the main point of our paper, we have not dedicated a figure to this here.  (We simply note that the offset between the two sets of symbols in Figure~\ref{compareWp} is a direct consequence of this luminosity dependence.)

Next, from each catalog we chose the red subset which had observed $g-i>2.35$, and measured $\xi(s)$ and $w_p(r_p)$ in it.  These red galaxies have comoving number densities which are about 20\% smaller than those of their parent volume limited catalogs, and they are also more strongly clustered.  
E.g., the red objects have a bias factor which is about $5/3$ times that of the blue objects (those which have $g-i<2.35$).  The reddest galaxies in a luminosity threshold sample are known to be more strongly clustered than the rest (e.g. Skibba \& Sheth 2009; Guo et al. 2013), so it is reassuring that, despite the relatively large errors in the photometry, this extremely simple color cut appears to have removed a physically different sample with a substantially smaller clustering signal.  These are the measurements which we use for our `conserved tracers' test of passive evolution.  

The squares in Figure~\ref{compareWp} show our measurements of $w_p$ for the red CMASS galaxies in our brightest and faintest volume limited catalogs, divided by a fiducial value $w_p^{\rm fid}$, for which we use the measurement for the full CMASS sample reported in Table~2 of Reid et al. (2014).  The rarer, more luminous galaxies are clearly more strongly clustered, being offset from $w_p^{\rm fid}$ by a scale independent multiplicative factor over all scales larger than a few Mpc.  

The triangles connected by a solid line show similar clustering measurements made in volume limited catalogs of the most luminous SDSS DR7 E+S0 galaxies, with the limits chosen to yield the same comoving number density as the CMASS red galaxies:  $z\le (0.2,0.24)$ and {\tt cmodel} $M_i \le (-22.77, -23.17)$.  These catalogs have comoving volumes that are about $7\times$ smaller than their CMASS counterparts, so the error bars on the SDSS measurements are correspondingly larger.  This shows that the more luminous E+S0s at $z\sim 0.2$ are slightly less clustered than their red CMASS counterparts at $z\sim 0.55$ whereas the less luminous galaxies are substantially less clustered.  We find, but do not show here, similar results using $\xi(s)$.  This is reassuring, in view of our comments earlier about systematics associated with the $w_p$ measurement (and which our use of ratios to present results mitigates).  Since the conserved tracer assumption predicts that the low redshift sample should be more strongly clustered, we conclude that our clustering measurements are inconsistent with pure passive evolution.  In fact, for the reasons given at the start of this section, the clustering measurements are inconsistent with any merger model which preserves the rank ordering in luminosity.  

\subsection{Systematics and an additional test}
In view of how very passive both $\phi(L)$ and $\phi(M_*)$ seem to be, it is prudent to consider how the clustering test may have gone wrong.

The completeness of the CMASS sample is still under investigation (e.g. Leauthaud et al. 2015 and our Figure~\ref{LFred} suggest this is of order of 80 percent).  However, because we have been careful to match comoving number densities in SDSS and CMASS, any incompleteness in CMASS would almost certainly mean that our current CMASS samples contain more lower luminosity galaxies than they should have (for a given number density cut).  Since lower luminosity galaxies are less strongly clustered, incompleteness in CMASS would mean that we have underestimated the clustering strength of the $z\sim 0.55$ population.  Thus, incompletenesses in the CMASS sample will only exacerbate the mismatch with the passive evolution prediction.  

Another possibility is that we are somehow underestimating the clustering signal of the low-$z$ sample.  For example, perhaps the $p$(type) weights we use are not sufficiently reliable, and yield a systematic underestimate of the clustering strength.  This may be:  the brightest SDSS$_{\rm CMASS}$ E+S0s in Figure~\ref{compareWp} have $w_p/w_p^{\rm fid}\approx 1.7$, and this is not very different from the value of $\sim 1.6$ in the bottom panel of Figure~\ref{xi79}, for which no $p$(type) weights were applied.  On the other hand, although we expect E+S0s to be more strongly clustered than the total, this is most dramatic at faint luminosities (e.g., because faint satellite galaxies in clusters tend to be early-type).  At the highest luminosities of interest here, most galaxies are E+S0s, so the expected difference is small.  E.g., Zehavi et al. (2011) suggest that this difference is less than 10 percent for the most luminous (lowest comoving number density) sample.  

Moreover, the similarity we see in the clustering strength of the rarest objects is consistent with the analysis of luminous red galaxies (LRGs) presented in Wake et al. (2008).  They find that the clustering of LRGs (approximately equivalent to our rarer more luminous sample) has evolved little between $z=0.55$ and $z=0.2$:  it has not increased.  They attribute this to mergers involving a small fraction of objects (our Appendix~B discusses a simple toy model which illustrates why the clustering signal decreases because of mergers).  
While this agreement may be reassuring, we note that their measurements of $\phi(L)$ were not as precise as ours -- the precision of the $\phi(M_*)$ measurements shown in the previous section leaves little room for mergers.  

\begin{figure}
 \centering
 \includegraphics[scale = .4]{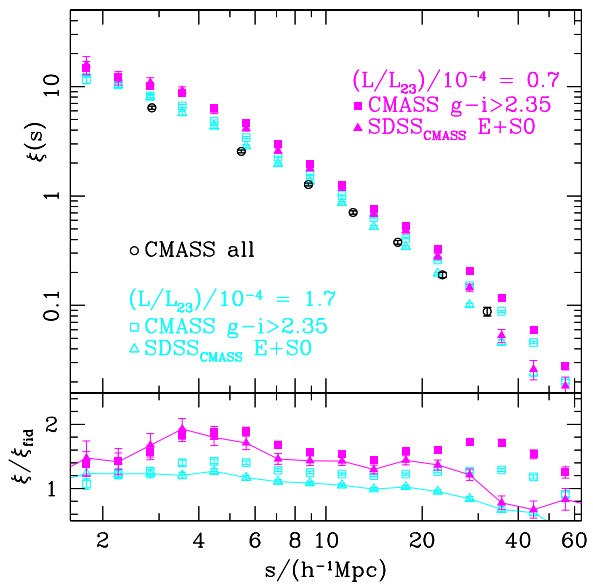}
 \caption{Redshift space correlation function of the full CMASS sample (open circles), and two subsamples of CMASS red galaxies chosen to have the luminosity densities indicated (squares), and SDSS$_{\rm CMASS}$ E+S0 galaxies with same luminosity density (triangles).  Bottom panel shows the ratio of these measurements to that in the full sample.  If the conserved tracer/passive-like hypothesis were correct, then the SDSS$_{\rm CMASS}$ galaxies would be about 15\% more strongly clustered than their CMASS counterparts.}
 \label{xiLpassive}
\end{figure}

A final possibility is that we are simply not making the appropriate comparison between the low and high redshift samples.  Motivated by the fact that the stellar mass of a galaxy which results from the merger of two passive galaxies will almost certainly be close to that of the sum of the masses of its progenitors, Tojeiro et al. (2012) advocate testing for passive evolution by using samples matched in (comoving) luminosity- or stellar-mass density, rather than in number density.  (A little thought shows that this is not as clean a test as advertised:  If CMASS galaxies merged with non-CMASS galaxies, then although the number density of CMASS galaxies is conserved, the luminosity- or stellar-mass density is not, so matching them -- rather than number density -- at different times is no longer appropriate.)  Since we have found that $\phi(L)$ and $\phi(M_*)$ are so close to passive, we do not expect $L$ or $M_*$- weighting each galaxy to make much of a difference.  Nevertheless, we have performed such a test using the {\tt cmodel} luminosities.  

The brighter and fainter samples studied previously have luminosity densities of 0.7 and $1.7\times 10^{-4}L_{23}/(h^{-1}{\rm Mpc})^3$, where $L_{23}$ is the luminosity associated with {\tt cmodel} $M_i=-23$.  (The mean luminosity in each sample is 1.45 and 1.05$L_{23}$)  We then measured the luminosity-weighted $\xi(s)$ and $w_p(r_p)$ in these matched CMASS and SDSS$_{\rm CMASS}$ samples (i.e., for the SDSS$_{\rm CMASS}$ sample, each galaxy was weighted by $p$(E)+$p$(S0) times $L/L_{23}$.  To make the point that both give similar results, we now show $\xi(s)$, since we previously showed $w_p$.  Note that the conserved tracer/passive evolution prediction for the redshift space ratio -- a growth of about 15\% between $z=0.55$ and $z=0.2$ -- is slightly smaller than for the real-space ratio (equation~\ref{xisratio}).  

Squares and triangles in Figure~\ref{xiLpassive} show results for our CMASS and SDSS$_{\rm CMASS}$ subsamples.  The open circles show $\xi(s)$ for the full CMASS sample taken from Table~2 in Reid et al. (2014; since these were not luminosity-weighted, they correspond to the open circles in Figure~\ref{xi79}).  The bottom panel shows the ratio of our luminosity-weighted measurements to the open circles.  As for $w_p$, this ratio is rather scale independent.  And, as before, the SDSS$_{\rm CMASS}$ signal does not exceed that for CMASS on scales larger than a few Mpc.  (The drop on large scales for the most luminous sample is most likely due to cosmic variance.)  Since the prediction was an increase of about 15\%, we conclude that the conserved tracer/passive evolution hypothesis is inconsistent with the results of this test as well.

Having argued that our clustering results seem to be robust, we now consider the possibility that systematic errors in our determination $L$ or $M_*$ are to blame.  We noted at the start of this section that because we match comoving densities we are immune to systematic problems in the models we use for converting from apparent magnitudes to luminosities ($k+e$ corrections) and from $L$ to $M_*$, provided these leave the rank ordering the same.  This same argument applies to systematics in the photometric reductions (e.g., Meert et al. 2014 showed that our Sersic reductions are slightly biased; also see D'Souza et al. 2015) if they leave the rank ordering of luminosities the same.  However, this is not quite the full story.  Suppose the systematic is different for CMASS galaxies than it is for SDSS$_{\rm CMASS}$ (e.g., because of surface-brightness effects, etc.), but it preserves the rank ordering in each.  Then, although our clustering measurements will not change (because the rank ordering is unchanged), our interpretation will, because $\phi(L)$ and $\phi(M_*)$ will no longer be consistent with passive evolution.  This might provide room for the mergers which our clustering results suggest have occured.  Although the tests in Meert et al. (2014) suggest that such a systematic is not present in {\tt PyMorph}, we believe our results motivate further testing, ideally by other groups with different analysis pipelines.

\section{Conclusions} 
We showed that our Sersic-based photometric reductions imply more luminous, massive galaxies in the CMASS sample (at $z\sim 0.55$) than when {\tt cmodel} photometry is used (Figure~\ref{pymorph}).  This difference is consistent with the effects of Sersic- rather than Petrosian or de~Vaucouleur-based photometry on the SDSS main galaxy sample at $z\sim 0.1$.  This implies a significant revision of the high mass end of the correlation between stellar and halo mass, and impacts the need for feedback processes operating at $z\sim 1$ (e.g. Kravtsov et al. 2014; Shankar et al. 2014).  

Inferences about the evolution of the luminosity and stellar mass functions, and hence of the $M_*-M_{\rm halo}$ relation, depend strongly on the assumed and uncertain $k+e$ corrections.  In most stellar population synthesis codes, these depend strongly on the age of the stellar population.  The models of M09 suggest that CMASS galaxies are about 2~Gyrs younger than Wake et al. (2006) assumed for LRGs at these same redshifts (Figure~\ref{keage}).  They also indicate that more luminous CMASS galaxies are older (although not as old as the LRGs were assumed to be).  This implies differential evolution across the population, and suggests that use of a single $k+e$ correction for all galaxies at the same $z$ will lead to biases, even if they are all early-type.  

To test the hypothesis that CMASS galaxies evolve passively to populate the high mass/luminosity end of the SDSS sample, we described how to use the M09 models to construct a mock CMASS sample from the SDSS sample.  We called the result the SDSS$_{\rm CMASS}$ sample; if passive evolution is accurate then the luminosity and stellar mass functions in it should be the same as in CMASS.  They are:  passive evolution is an even better approximation if we compare red galaxies in CMASS (observed $g-i>2.35$) with E+S0s in SDSS$_{\rm CMASS}$ (Figure~\ref{LFred} and~\ref{MFred}).  


If the CMASS and SDSS$_{\rm CMASS}$ galaxies are indeed related by simple passive evolution, then the conservation of comoving number density implies that the SDSS$_{\rm CMASS}$ population should be about 20 percent more strongly clustered than their counterparts in CMASS.  To test this prediction, we matched samples in comoving number density ($\sim 2\times 10^{-4} h^3$~Mpc$^{-3}$) and type ($g-i>2.35$ in CMASS and morphological type E+S0 in SDSS$_{\rm CMASS}$), and measured the clustering signals in each, finding that SDSS$_{\rm CMASS}$ sample is less clustered than its CMASS counterpart (Figure~\ref{compareWp}).  I.e., the low redshift clustering signal lies well below the passive evolution/conserved tracer prediction.  

We repeated the test using more luminous, less abundant galaxies (comoving densities $\sim 0.5\times 10^{-4} h^3$~Mpc$^{-3}$).  In both CMASS and SDSS$_{\rm CMASS}$, the more luminous objects are more strongly clustered.  In addition, at fixed luminosity, the redder CMASS objects are more strongly clustered:  simply requiring $g-i>2.35$ removes a physically different, substantially less clustered population of objects from the CMASS sample.  
(Both these findings extend recent analyses of trends with luminosity and color in the CMASS sample by Guo et al. (2013) to higher luminosities and masses.)  However, although the differences between the CMASS and SDSS$_{\rm CMASS}$ clustering strengths decrease for the rarer objects, in no case was the SDSS$_{\rm CMASS}$ sample more strongly clustered than the corresponding CMASS sample (Figure~\ref{compareWp}):  pure passive evolution is always a bad approximation.  

The nature of the clustering test means that our measurements are immune to systematic biases in the models we use for converting from apparent magnitudes to luminosities ($k+e$ corrections) and from $L$ to $M_*$, provided these leave the rank ordering of $L$ or $M_*$ the same.  They are also inconsistent with any merger model which preserves the rank ordering in luminosity.  
Incompleteness in the CMASS sample will likely amplify rather than mitigate this discrepancy.  Matching the CMASS and SDSS$_{\rm CMASS}$ samples in comoving luminosity (rather than number) density, and weighting each galaxy by its luminosity when measuring the clustering signal leads to a similar conclusion:  since the low redshift sample is never more strongly clustered than the high redshift sample (Figure~\ref{xiLpassive}), pure passive evolution is inconsistent with our measurements.  

While this is in conflict with our findings based on the luminosity and stellar mass functions, it is consistent with analyses of the abundance and clustering of LRGs in Wake et al. (2008).  Indeed, our finding that clustering rules out passive evolution means that the merger models in Wake et al. (2008) for the evolution of LRGs are also likely to be relevant for CMASS.  There also exists a large body of work which suggests -- primarily on the basis of number counts alone -- that there must have been of order 0.1 to 0.2~dex mass growth via mergers over the redshifts and masses of interest here (see e.g. Marchesini et al. 2014; Ownsworth et al. 2014 and references therein).  Although we outlined a simple toy model for how abundance {\em and} clustering measurements constrain merger models (Appendix~B), the exquisitely passive nature of $\phi(M_*)$ which we have found is a puzzle which we hope will spur further work.  In particular, the larger CMASS sample size has allowed a more precise determination of the luminosity and stellar mass functions than was possible in Wake et al. (2006); this means that, if the M09 models -- on which our $k+e$ corrections and hence our evidence for passive evolution -- are correct, then there is substantially less room for mergers to play a role.  

The discrepancy between the abundances which are consistent with passive evolution and the clustering measurements which are not, might be alleviated if our photometric reductions suffer from a systematic bias which affects $z\sim 0.2$ galaxies differently from those at $z\sim 0.6$.  While the tests in Meert et al. (2014) have failed to uncover such a systematic, we believe this tension motivates further work -- perhaps by other groups with different photometric analysis pipelines -- along these lines.  

Before closing, we note that the passive evolution/conserved tracers prediction is robust to currently acceptable changes in the $\Lambda$CDM model parameters.  Hence, if both the stellar population synthesis models and our clustering analyses are correct -- passive evolution of the stellar population and weaker than expected clustering at late times -- then the overprediction of the clustering signal may be pointing to new physics.  E.g. faster than expected expansion between $z=0.55$ and the present would lead to a freezing-out of structure formation, thus reducing the tension between the SSP models and the lack of evolution in the clustering signal.  However, this is a rather radical solution -- one to which we are reluctant to turn -- given current uncertainties on the models and the modest comoving volume currently available to perform the clustering test.  Nevertheless, as larger samples over larger comoving volumes become available, we believe that the combination of SSP and clustering analyses we have used here will yield interesting results.

\subsection*{Acknowledgements}
We are grateful to A. Montero-Dorta for helpful discussions about his work, and the staff of the LYTE center for its hospitality when this work was completed.

\appendix
\section{Photometric errors}\label{errors}
In this Appendix we address the question of how photometric errors affect our results.  We have used our SDSS$_{\rm CMASS}$ mock catalogs to do this as follows.  

The dotted curves in the bottom panel of Figure~\ref{LFred} show the luminosity function in which we use the broadened photometric errors to mimic CMASS conditions; the dashed curves in our SDSS$_{\rm CMASS}$ sample use the unbroadened (i.e. original, SDSS) photometry of the SDSS$_{\rm CMASS}$ objects.  The difference between the dotted and dashed curves is an indicator of how photometric errors affect $\phi(L)$.  These matter little at the bright end, as one might expect.  However, they make a substantial difference at the faint end of the {\tt PyMorph} counts, even though they mattered little for the {\tt cmodel} $i$-band.  This apparent dependence on photometric reduction arises because the sample is selected on the basis of observed $i$-band {\tt cmodel} photometry:  i.e., after being broadened by the errors.  Therefore, it is only the (rather narrow) redshift distribution which transforms the sharp cut in apparent magnitude into a more gradual (but still quite sharp) cut in absolute magnitude.  The differences between {\tt cmodel} and {\tt PyMorph} photometry transform the sharp {\tt cmodel} cut into something fuzzier; this makes the red dashed curve decline more gradually than the green.  The errors then broaden this further.  

\begin{figure}
 \centering
 \includegraphics[scale = .43]{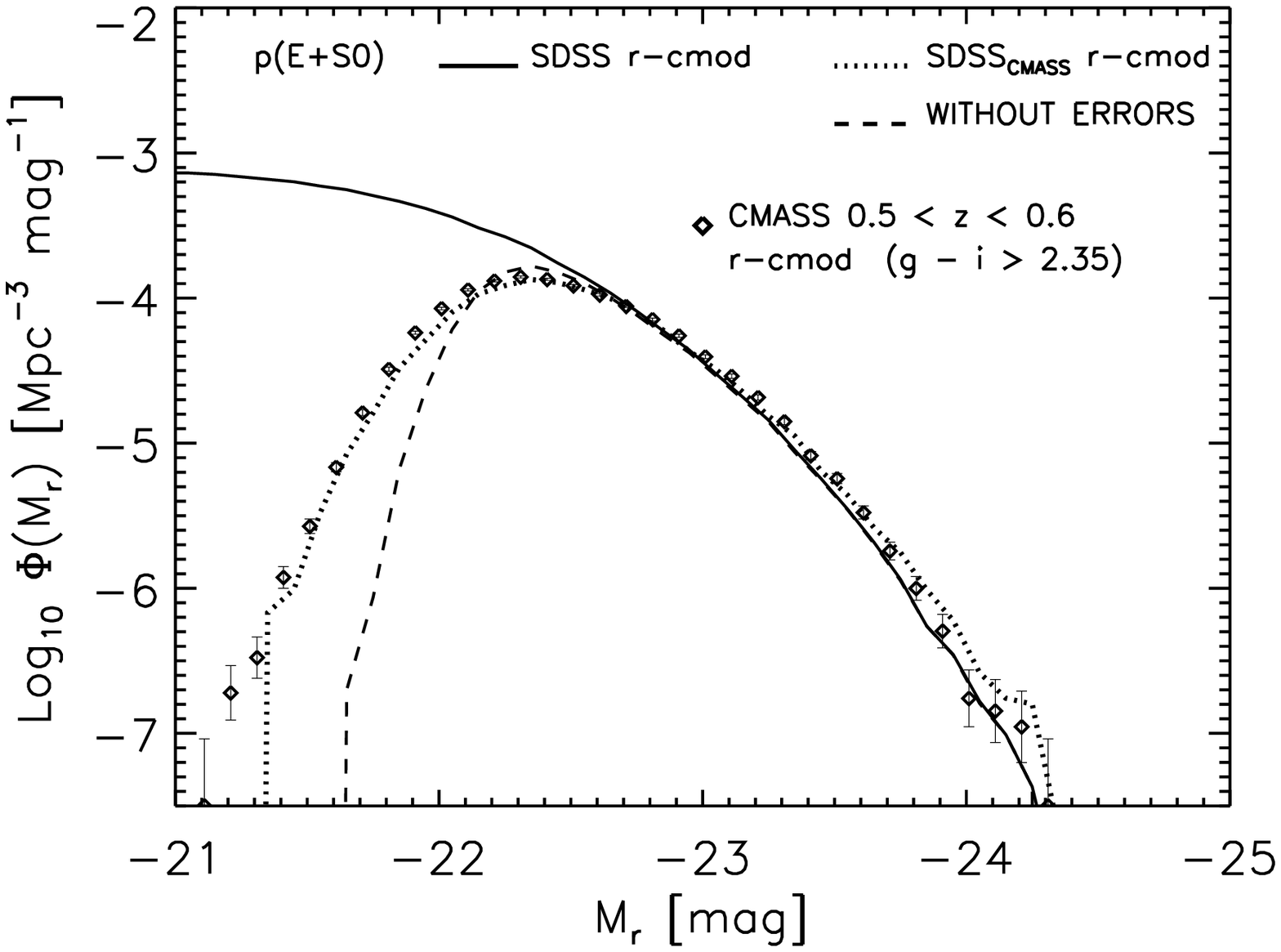}
 \caption{Luminosity functions from {\tt cmodel} $r$-band photometry.  Symbols show our measurements in CMASS; dotted curves show the SDSS$_{\rm CMASS}$ mock catalog; solid curves show the SDSS main galaxy sample.  Dashed curves show the distribution in SDSS$_{\rm CMASS}$ if the photometric errors on the fainter (because they are more distant) SDSS$_{\rm CMASS}$ galaxies were the same size as for the brighter SDSS galaxies.  These make a bigger difference at the faint end than the bright.}
 \label{LFri}
\end{figure}

This effect -- that errors matter more at the faint end than the bright -- applies whenever one is working with photometry which is not the one in which the sample was selected (in this case, the {\tt cmodel} $i$-band).  To further make this point, Figure~\ref{LFri} shows the luminosity function in the {\tt cmodel} $r$-band.  Notice that the dashed curve is very different from the dotted one (both made using our SDSS$_{\rm CMASS}$ mock catalogs) at the faint end, for the reasons discussed above (i.e., the sample was defined using $i$-, not $r$-band photometry).  On the other hand, the dotted curves (from SDSS$_{\rm CMASS}$) and symbols (from CMASS) are in excellent agreement, just as they were in the bottom panel of Figure~\ref{LFred}.  This agreement shows that {\tt cmodel} $r$-band photometry leads to the same conclusion as the {\tt cmodel} $i$-band photometry -- passive evolution is an excellent description if the M09 $k+e$ corrections are realistic -- so we believe we have correctly modelled the errors. 

Although we do not show it here, analysis of {\tt model} photometry in the $r$- and $i$-bands, as well as the higher redshift range $0.6\le z\le 0.7$ yields similar results.

\section{A simple merger model}\label{sec:toy}
The main text argued that, if the CMASS galaxies are to evolve into the most massive SDSS galaxies, then this must have happened via mergers which shuffle the rank ordering of stellar mass or luminosity between $z\sim 0.55$ and $z\sim 0.2$.  Since the average mass increase is small, major mergers across the majority of the CMASS sample are ruled out.  However, major mergers for a small fraction of the sample may be permitted.  Since such mergers will not preserve the $z=0.55$ rank-ordering in stellar mass, it is interesting to ask if such a major merger model can be consistent with both the number counts and the clustering.

\subsection{Constraints from the evolution of the abundances}\label{toyphi}
To illustrate the argument, we suppose that a randomly chosen fraction $f$ of the CMASS sample evolves passively from $z\sim 0.55$ to the present time.  The remainder undergoes equal mass major mergers with other CMASS galaxies, such that $\phi(M_*)$ for these objects is decreased in abundance by a factor of two and shifted towards masses which are larger by factor of two.  (We will argue shortly that it may be better to think of the `passive' subset as having been involved in minor mergers with non-CMASS galaxies while preserving the rank ordering in mass.)  


Decreasing $f$ corresponds to increasing the fraction of objects which underwent major mergers, thus increasing the predicted large $M_*$ tail.  (E.g., if $f\to 0$ then the entire population would shift to masses larger by a factor of two.)  The fact that the mismatch at large $M_*$ shown in the main text is only of order 0.05~dex (or smaller) constrains $f \sim 0.9$, thus ruling out the need for major mergers across all but a small fraction of the sample.

\subsection{Evolution of clustering in the simple merger model}\label{toyxi}
To see if such a model is also consistent with the clustering, suppose that $b_p$ denotes the passively evolved bias of the fraction $f$ of the high-$z$ population which evolved via minor mergers that preserved the comoving number density and rank ordering in mass.  It is reasonable to suppose that the population which underwent major mergers had a different bias:  we use $b_m$ to denote the bias of this population had it too evolved passively, and we expect $b_m\ge b_p$ (because larger clustering signal means more nearby neighbours, and so increased likelihood of merging).  We will use $B$ to denote the net bias of the combined population had it evolved passively.  

Since the major mergers are assumed to have been with other members of the CMASS sample, they reduce the number density by a factor of $f + (1-f)/2 = (1+f)/2$:  the luminosity or stellar mass function constrains $f\approx 0.8$, so the post-merger number density is $0.9\times$ the original one.  However, major mergers also reduce the net bias of the combined population.  To see why, note that, absent mergers, the number of pairs at separation $r$ satisfies 
\begin{eqnarray}
 &&(1-f)^2(1 + b_2^2\xi) + f^2(1 + b_1^2\xi) + 2f(1-f)(1 + b_1b_2\xi)\nonumber\\
 &&     = [(1-f)^2 + f^2 + 2f(1-f)]\,(1 + B^2\xi),
 \label{xi2pops}
\end{eqnarray}
with $(b_1,b_2)=(b_p,b_m)$.  Now, if the population with $b_m$ merges with other members of $b_m$, reducing their numbers by half, then equation~(\ref{xi2pops}) should be modified to read
\begin{eqnarray}
 &&(1-f)^2(1 + b_m^2\xi)/4 + f^2(1 + b_p^2\xi) + f(1-f)(1 + b_pb_m\xi)\nonumber\\
 &&     = [(1-f)^2/4 + f^2 + f(1-f)]\,(1 + B_{\rm mod}^2\xi) \\
 &&     = [(1+f)/2]^2\,(1 + B_{\rm mod}^2\xi), \nonumber
\end{eqnarray}
where $B_{\rm mod}$ denotes the bias of the population which results.  Thus, 
\begin{equation}
  B_{\rm mod}/B = (1 + f b_p/B)/(1 + f).
\end{equation}
When combined with $(1-f)(b_m/B) = 1 - f\,(b_p/B)$, this implies that 
\begin{eqnarray}
 f\,\frac{b_p}{B} &=& (B_{\rm mod}/B) (1+f) - 1 \qquad {\rm and}\\
 \frac{b_m}{B} 
        &=& \frac{2 - (1+f)(B_{\rm mod}/B)}{1-f}.
\end{eqnarray}
Since $B_{\rm mod}/B$ is observable, it, along with the constraint on $f$ which one obtains from $\phi(M_*)$, constrains $b_p$ and $b_m$.  Note in particular that if $B_{\rm mod}<B$ then $b_p<b_m$.  

Most galaxy populations have bias factors which are not too different from unity.  Thus, if $b_m$ is too large or $b_p$ too small, we might conclude that major mergers are disfavored.  In our case, the stellar mass function suggests $f=4/5$ (or even closer to unity), and the clustering suggests $B_{\rm mod}/B = 8/9$, so $b_p/B = 3/4$ and $b_m/B = 2$.  Since $B$ is slightly larger than 2, this requires $b_p\sim 1.5$ at $z=0.2$, and so $\sim 1.6$ at $z=0.55$.  This is typical of the halos at the low mass end of the CMASS sample.  On the other hand, $b_m\sim 4$ corresponds to $\sim 4.6$ at $z=0.55$, which is of the order expected for the most massive halos.  Wake et al. (2008) explore more complicated models than ours to interpret their (qualitatively similar) clustering measurements of LRGs over the same redshift range.  Their Figures~13 and~15 show that they believe the mergers are associated with halos more massive than $10^{14}h^{-1}M_\odot$, which is consistent with our crude $b_m$ estimate.  

Thus, it appears that our clustering measurements are consistent with plausible merger models in which the number density of tracers is not conserved:  a small fraction of the CMASS galaxies experienced major mergers (doubled their mass), and most of the other CMASS galaxies experienced minor mergers which 
preserved the comoving number density and rank ordering in mass.  Exploring other models in which the major mergers do not exactly double the mass, and/or a fraction of the CMASS galaxies experience no mergers, is beyond the scope of this paper.


\begin{thebibliography}{}
 \bibitem{} 
  Abazajian, et al., 2009, ApJS, 182, 543
 \bibitem{}
  Anderson L., Aubourg E., Bailey S., et al., 2012, MNRAS, 427, 3435
 \bibitem{}
  Banerji M., Ferreras I., Abdalla F. B., Hewett P. \& Lahav O., 2010, MNRAS, 402, 2264 
 \bibitem{} 
  Bell E.~F., McIntosh D.~H., Katz N. \& Weinberg M.~D., 2003, ApJS, 149, 289
 \bibitem{}
  Bernardi M., Shankar F., Hyde J. B., Mei S., Marulli F. \& Sheth R. K., 2010, MNRAS, 404, 2087 (B10)
 \bibitem{} 
  Bernardi M., Roche N., Shankar F. \& Sheth R. K., 2011a, MNRAS, 412, L6
 \bibitem{} Bernardi M., Roche N., Shankar F. \& Sheth R. K., 2011b, MNRAS, 412, 684
 \bibitem{}
  Bernardi M., Meert A., Sheth R. K., Vikram V., Huertas-Company M., Mei S. \& Shankar F., 2013, MNRAS, 436, 697
 \bibitem{}
  Bernardi M., Meert A., Vikram V., Huertas-Company M., Mei S., Shankar F. \& Sheth R. K., 2014, MNRAS, 443, 874
 \bibitem{}
  Bruzual G. \& Charlot S., 2003, MNRAS, 344, 1000
 \bibitem{}
  Cappellari M., McDermid R. M., Alatalo K., et al., 2013, MNRAS, 432, 1862 
 \bibitem{}
  Charlot S. \& Bruzual G., 2007, Currently unpublished. An updated Bruzual G. \& Charlot S. 2003 model (CB07)
 \bibitem{}
  Cimatti A., Daddi E. \& Renzini, A. 2006, A\&A, 453, L29 
 \bibitem{}
  Cooray A. \& Sheth R. K., 2002, Phys Rep, 1
 \bibitem{} D'Souza R., Vegetti S. \& Kauffmann G., 2015, MNRAS, in press (arXiv:1509.07418)
 \bibitem{}
  Guo H., Zehavi I., Zheng Z., et al., 2013, ApJ, 767, 122
 \bibitem{} 
  Huertas-Company M., Aguerri J. A. L, Bernardi M., Mei S. \& S{\'a}nchez Almeida J., 2011, A\&A, 525, 157
 \bibitem{}
  Kaiser N., 1987, MNRAS, 227, 1
 \bibitem{}
  Kravtsov A., Vikhlinin A. \& Meshscheryakov A., 2014, Apj, submitted (arXiv:1401.7329)  
 \bibitem{}
  Leauthaud A., et al., 2015, MNRAS, submitted
 \bibitem{}
  Maraston C., Str{\"o}mb{\"a}ck G., Thomas D., Wake D. A. \& Nichol R. C., 2009, MNRAS, 394, L107 (M09)
 \bibitem{}
  Maraston C., Pforr J., Henriques B. M., et al., 2013, MNRAS, 435, 2764 (M13)
 \bibitem{}
  Marchesini D., Muzzin A., Stefanon M., et al., 2014, ApJ, 794, 65
 \bibitem{}
  Meert A., Vikram V. \& Bernardi M., 2013, MNRAS, 433, 1344
 \bibitem{}
  Meert A., Vikram V. \& Bernardi M., 2015, MNRAS, 446, 3943
 \bibitem{}
  Mo H. J. \& White S. D. M., 1996, MNRAS, 282, 347
 \bibitem{}
  Montero-Dorta A. D., Bolton A. S., Brownstein J. R., et al., 2015, MNRAS, submitted (arXiv:1410.5854) 
 \bibitem{}
  Nusser A., Davis M., 1994, ApJ, 421, L1
 \bibitem{}
  Nuza S. E., Sanchez A. G., Prada F., et al., 2013, MNRAS, 432, 743 
 \bibitem{}
  Ownsworth J. R., Conselice C. J., Mortlock A., Hartley W. G., Almaini O., Duncan K., Mundy C. J., 2014, MNRAS 445, 2198
 \bibitem{} 
  Reid B. A., Seo H., Leauthaud A., Tinker J. L. \& White M., 2014, MNRAS, 444, 476
 \bibitem{}
  S{\'e}rsic J. L., 1963, Bol. Asociacion Argentina Astron. La Plata Argentina, 6, 41
 \bibitem{}
   Shankar F., Guo H., Bouillot V., et al., 2014, ApJL, 797, 27
 \bibitem{}
  Skibba R. A., Sheth R. K., 2009, MNRAS, 392, 1080
 \bibitem{}	
  Tojeiro R., Percival W. J., Brinkmann J., et al., 2012, MNRAS, 424, 2339
 \bibitem{}
  Wake D. A., Nichol R. C., Eisenstein D. J., et al., 2006, MNRAS, 372, 537
 \bibitem{}
  Wake D. A., Sheth R. K., Nichol R. C., et al., 2008, MNRAS, 387, 1045
\end{thebibliography}
\end{document}